\documentclass[%
 reprint,
superscriptaddress,
 amsmath,amssymb,
 aps,
pre,
]{revtex4-2}

\usepackage{graphicx}
\usepackage{dcolumn}
\usepackage{bm}

\usepackage{wrapfig}
\usepackage{enumitem}
\usepackage{amsmath,stackengine}
\usepackage{amssymb}
\usepackage{marginnote}
\usepackage[dvipsnames]{xcolor}
\usepackage[colorlinks, allcolors=NavyBlue]{hyperref}
\usepackage{url}
\usepackage{esint}
\usepackage{multirow}
\usepackage{tikz}
\usetikzlibrary{arrows.meta}

\usepackage{mathtools}
\usepackage{listings}
\usepackage{siunitx}
\usepackage{pgfplots}
\pgfplotsset{compat=1.9, , every axis/.append style={font=\scriptsize}}
\usepackage{xspace}
\usepackage{orcidlink}
\usepackage[caption=false]{subfig}

\newcommand{\Cel}{\mathfrak{C}^{\text{el}}}

\renewcommand{\phi}{\varphi}

\newcommand{\brak}[1]{\langle #1 \rangle}

\begin{document}

\title{Final state sensitivity and fractal basin boundaries from coupled Chialvo neurons}

\author{Bennett Lamb}
\affiliation{Thomas Jefferson National Accelerator Facility, Newport News, Virginia 23606, USA}
\author{Brandon B. Le}
\email[Contact author: ]{sxh3qf@virginia.edu}
\affiliation{
Department of Physics, University of Virginia, Charlottesville, Virginia 22904, USA
}

\date{\today}

\begin{abstract}
We investigate and quantify the basin geometry and extreme final state uncertainty of two identical electrically asymmetrically coupled Chialvo neurons. The system's diverse behaviors are presented, along with the mathematical reasoning behind its chaotic and nonchaotic dynamics as determined by the structure of the coupled equations. The system is found to be multistable with two qualitatively different attractors. Although each neuron is individually nonchaotic, the chaotic basin takes up the vast majority of the coupled system's state space, but the nonchaotic basin stretches to infinity due to chance synchronization. The boundary between the basins is found to be fractal, leading to extreme final state sensitivity. This uncertainty and its potential effect on the synchronization of biological neurons may have implications for understanding neuronal biology.
\end{abstract}

\maketitle


\section{Introduction}
\label{sec:intro}
With only three parameters and two state variables, the Chialvo map~\cite{chialvo} is well established to be able to accurately mimic the behavior of biological neurons, and its simplicity allows for thorough investigations into its state space, dynamics, and geometries. Since the pioneering work of Hodgkin and Huxley~\cite{hh}, many continuous-time neuron models have been developed, including Izhikevich~\cite{izhikevich-article}, Hindmarsh-Rose~\cite{hindmarsh-rose}, and FitzHugh-Nagumo neurons~\cite{fh}. However, discrete-time systems can be advantageous for neuronal modeling due to their simultaneous simplicity and ability to exhibit a wide range of behaviors. Discrete-time neuron models have been developed through various mechanisms: some maps, like the Chialvo map~\cite{chialvo} and the Rulkov map~\cite{rulkov} are discrete-time by design, while others have come from continuous-time counterparts. For example, the Izhikevich, Hindmarsh-Rose, and FitzHugh-Nagumo neurons have all been adapted to discrete-time maps~\cite{discrete-adaptations,hr-discrete,fh-discrete}. 

These neuron models continue to be studied extensively for their dynamics. In recent years, the dynamics of the Hindmarsh-Rose~\cite{hr-dynamics,hr-dynamics-2,hr-dynamics-3}, FitzHugh-Nagumo~\cite{fh-dynamics,fh-dynamics-2,fh-dynamics-3,yanagita-fh}, Izhikevich~\cite{izh-dynamics,izh-dynamics-2,izh-dynamics-3}, and Rulkov neurons~\cite{ring,n-dim} have been the subject of continued investigation. In Ref.~\cite{yanagita-fh}, the authors characterize the chaotic dynamics of a pair of Fitzhugh-Nagumo neurons, focusing on similar quantitative measures of uncertainty and basin analysis. Additionally, a ``vulnerability of synchronization'' was found in simulated neural circuits~\cite{medeiros_vulnerability_of_sync}, where it was found that a perturbation sent an orbit to different attractors depending on when the perturbation was applied. Clearly, oscillatory dynamical systems are an active area of investigation.

The motivation of this paper is to extend the research on final state sensitivity and basin geometry of discrete-time neuronal systems. Multiple continuous-time systems have shown final state uncertainty, including a network of theta neurons~\cite{so}, an adaptive synapse-based neuron model with heterogeneous attractors~\cite{bao-neuron}, and a network of Izhikevich neurons~\cite{aristides}. However, discrete-time research is more limited. Recently, two electrically and asymmetrically coupled Rulkov neurons were also found to exhibit final state uncertainty~\cite{le}, and the ubiquity of uncertainty in discrete-time neuron systems was demonstrated~\cite{ubiquity}. This paper seeks to solidify the claim that final state uncertainty is ubiquitous in discrete-time neuron systems, as well as to provide a more in-depth look at the basin geometry and uncertainty properties of electrically coupled Chialvo neurons.

Much work has been done to study the synchronization of coupled Chialvo neurons, although most rely on the stochastic model, where noise is introduced. Using this model, it has been shown that synchronization can be increased by random rewiring in a network of connected neurons~\cite{jampa}, by using excitatory connections rather than inhibitory connections between the neurons~\cite{chialvo-network,used1}, or by increasing the noise intensity and the coupling strength~\cite{two-chialvo}. Most recently, L\`{e}vy noise has been introduced to the system, and its impact on spiral wave and target wave patterns was investigated~\cite{kolesnikov,swetha}. More pertinent to our investigations, a ``sychronization malleability'' has been demonstrated in a large network of Chialvo neurons, where changes in the ordering of neural inputs results in different final states~\cite{sync-malleability}. In a large network of Chialvo neurons, it was shown that a small sub-population can have great influence on the whole network depending on the nature of the coupling~\cite{kamal-chialvo}. This demonstrates the sensitivity of the Chialvo model to perturbations in general, and we seek to further quantify this sensitivity in a more simplified model.

There has also been previous work on memristor-coupled Chialvo neurons. It has been observed that coherent behavior can appear and disappear as a result of coupling strength or coupling mechanism (e.g., electrical versus chemical coupling), where stronger electrical coupling leads to a more coherent system~\cite{flux-coupled-chialvo,chialvo-memristor-2}. Furthermore, for certain coupling strengths and parameter values, the system can exhibit transient synchronization~\cite{two-chialvo,chialvo-memristor} or hyperchaos~\cite{qin}. Recently, two distinct discrete-time neurons (Chialvo and Rulkov) were coupled via a memristor, and its synchronization was studied~\cite{chialvo-rulkov-memristor}. In another memristively-coupled heterogeneous system consisting of Chialvo and Izhikevich neurons, it was shown that the system was multistable and that each attractor could be reached by changing the initial state of the memristor~\cite{yan-heterochialvo}.

There have been investigations into the fractal nature of the parameter space. By varying the parameters $a$ and $b$ the parameter space has shown Neimark-Sacker bifurcation, Arnold tongues, shrimps, ``eyes of chaos"~\cite{eyes-of-chaos}, quasiperiodicity, pseudofractals, and hyperchaos~\cite{chialvo-memristor-3,wang-chialvo}. This has been shown for individual neurons and systems of two or more coupled neurons. These systems have also been shown to exhibit various routes to chaos, including period-doubling cascades and antimonotonicity~\cite{flux-coupled-chialvo-2}. 

The parameter space is well-documented for its fractal nature. Contrarily, this paper focuses on one particular choice of parameters to demonstrate that when each neuron is individually nonchaotic, the coupled map can still display chaotic behavior, and its state space can contain fractal basin boundaries. Moreover, through the previous studies, there is comparatively little discussion of final state sensitivity~\cite{grebogi-final-state}. In the original work by Chialvo~\cite{chialvo}, there is a short discussion of the different sub- and suprathreshold responses resulting from small gradations in initial conditions, and although this demonstrates the sensitive behavior of this map, we seek more quantitative results of its final state sensitivity, as well as a clear depiction of the different basins of attraction.

As a dedicated discrete-time system, the Chialvo model exhibits both periodic and aperiodic solutions through the variation of initial state and neuronal constants~\cite{chialvo}. This system is found to have two attractors, exhibiting multistability. Multistability is a rich concept in the study of a wide range of dynamical systems, through physics, biology, chemistry, and others, and controlling systems with multistability could have widespread effects~\cite{pisarchik_control_of_multistability}. However, the control of multistable systems is often very difficult. In a multistable system with final-state sensitivity, the basins and basin boundaries can exhibit various properties. For example, the boundaries can be fractal, or the basins themselves can be riddled~\cite{ott-riddled,alexander} or Wada~\cite{nusse}. These basins pose a significant obstruction to our ability to predict the final state of the model. This unpredictability and its effect on synchronization can have biological consequences---in the human brain, the synchronous firing of networks of neurons is necessary for many neural processes~\cite{fell-coupling,fell-phase-synchronization}, and asynchronous firing can lead to pathogenesis. While this behavior involves the interactions of large-scale neuron networks, the connection between these small-scale models and the large-scale networks may provide avenues for further research. With these goals, we seek to classify and quantify the uncertainty exhibited by the coupled Chialvo system. 

This paper is organized as follows. Sec.~\ref{sec:model} provides an investigation into the model itself and its dynamics. We then look deeper into the structure of the basins of attraction and analyze them through basin stability, basin classification, uncertainty exponent, and basin entropy methods in Sec.~\ref{sec:basin analysis}. The results are summarized, biological consequences are discussed, and future study is suggested in Sec.~\ref{sec:conclusions}. 

\section{The model and behavior}\label{sec:model}

In order to provide context for the analysis of the system's state space, we will first look into the different types of behavior that the system exhibits. The individual neuron exhibits behaviors that are echoed in the coupled system, but the latter is built on a more complex mathematical framework and thus has more complex dynamics. As such, understanding the individual system can provide valuable context to the rest of the investigation.

\subsection{The individual neuron}\label{sec:individual neuron}

The individual Chialvo neuron is a two-dimensional map of the form
\begin{align}
    \begin{split}\label{individual equations}
    \begin{cases}
        x_{n+1} &= x_n^2\exp(y_n-x_n)+I \\
        y_{n+1} &= ay_n - bx_n + c
    \end{cases}
    \end{split}
\end{align}
where $a,b,$ and $c$ are positive parameters and $I$ models the injection of current. $x$ represents the activation or voltage variable, while $y$ is the recovery variable. This map exhibits many behaviors typical of a neuron, including periodic, chaotic bursting, and quiescent solutions~\cite{chialvo}. 

\begin{figure*}
    \centering
    \subfloat[\label{cobweb1}]{
        \begin{tikzpicture}
            \begin{axis}[
                axis lines=middle,
                xmin=0, xmax=6,
                ymin=0, ymax=6,
                xlabel={$x_n$},
                ylabel={$x_{n+1}$},
                label style = {font=\large},
                xlabel style={below right},
                ylabel style={above left},
                width=8cm,
                height=8cm,
                samples=200,
                domain=0:6,
                legend pos=north east,
                ]
                \addplot[black, thick, dashed] {x};
        
                \addplot[
                    mark=*,
                    only marks,
                    mark size=2pt,
                    color=green
                ] coordinates {(0.44, 0.44)};

                \addplot[
                    mark=*,
                    only marks,
                    mark size=2pt,
                    color=red
                ] coordinates {(0.04, 0.04)};
        
                \addplot[blue, thick] {x^2 * exp(2.22 - x) + 0.04};
        
                \pgfmathsetmacro{\xstart}{0.44}
                \pgfmathsetmacro{\Nsteps}{1}
                \pgfmathsetmacro{\x}{\xstart}
                \foreach \i in {1,...,\Nsteps} {
                    \pgfmathsetmacro{\y}{\x^2 * exp(2.22 - \x) + 0.04}
                    \addplot[blue, thick, densely dashed] coordinates {(\x,\x) (\x,\y)};
                    \addplot[blue, thick, densely dashed] coordinates {(\x,\y) (\y,\y)};
                    \global\let\x\y
                    }
        
                \addplot[red, thick] {x^2 * exp(1.51 - x) + 0.04};
        
                \pgfmathsetmacro{\xstart}{1.19}
                \pgfmathsetmacro{\Nsteps}{1}
                \pgfmathsetmacro{\x}{\xstart}
                \foreach \i in {1,...,\Nsteps} {
                    \pgfmathsetmacro{\y}{\x^2 * exp(1.51 - \x) + 0.04}
                    \addplot[red, thick, densely dashed] coordinates {(\x,\x) (\x,\y)};
                    \addplot[red, thick, densely dashed] coordinates {(\x,\y) (\y,\y)};
                    \global\let\x\y
                    }

                \addplot[orange, thick] {x^2 * exp(-0.86 - x) + 0.04};
        
                \pgfmathsetmacro{\xstart}{2.00}
                \pgfmathsetmacro{\Nsteps}{1}
                \pgfmathsetmacro{\x}{\xstart}
                \foreach \i in {1,...,\Nsteps} {
                    \pgfmathsetmacro{\y}{\x^2 * exp(-0.86 - \x) + 0.04}
                    \addplot[orange, thick, densely dashed] coordinates {(\x,\x) (\x,\y)};
                    \addplot[orange, thick, densely dashed] coordinates {(\x,\y) (\y,\y)};
                    \global\let\x\y
                    }
        
                \addplot[purple, thick] {x^2 * exp(-5 - x) + 0.04};
        
                \pgfmathsetmacro{\xstart}{0.27}
                \pgfmathsetmacro{\Nsteps}{1}
                \pgfmathsetmacro{\x}{\xstart}
                \foreach \i in {1,...,\Nsteps} {
                    \pgfmathsetmacro{\y}{\x^2 * exp(-5 - \x) + 0.04}
                    \addplot[purple, thick, densely dashed] coordinates {(\x,\x) (\x,\y)};
                    \addplot[purple, thick, densely dashed] coordinates {(\x,\y) (\y,\y)};
                    \global\let\x\y
                    }
        
            \end{axis}
        \end{tikzpicture}
    }
    \hfill
    \subfloat[\label{cobweb2}]{
        \begin{tikzpicture}
    \begin{axis}[
        axis lines=middle,
        xmin=0, xmax=6,
        ymin=0, ymax=6,
        xlabel={$x_n$},
        ylabel={$x_{n+1}$},
        label style = {font=\large},
        xlabel style={below right},
        ylabel style={above left},
        width=8cm,
        height=8cm,
        samples=200,
        domain=0:6,
        legend pos=north east,
        ]
        \addplot[black, thick, dashed] {x};

        \addplot[
            mark=*,
            only marks,
            mark size=2pt,
            color=green
        ] coordinates {(0.41, 0.41)};
                
        \addplot[
            mark=*,
            only marks,
            mark size=2pt,
            color=red
        ] coordinates {(1.03, 1.03)};
        
        \addplot[blue, thick] {x^2 * exp(2.34 - x) + 0.03};

        \pgfmathsetmacro{\xstart}{0.41}
        \pgfmathsetmacro{\Nsteps}{1}
        \pgfmathsetmacro{\x}{\xstart}
        \foreach \i in {1,...,\Nsteps} {
            \pgfmathsetmacro{\y}{\x^2 * exp(2.34 - \x) + 0.03}
            \addplot[blue, thick, densely dashed] coordinates {(\x,\x) (\x,\y)};
            \addplot[blue, thick, densely dashed] coordinates {(\x,\y) (\y,\y)};
            \global\let\x\y
            }

        \addplot[red, thick] {x^2 * exp(2.29 - x) + 0.03};

        \pgfmathsetmacro{\xstart}{1.19}
        \pgfmathsetmacro{\Nsteps}{1}
        \pgfmathsetmacro{\x}{\xstart}
        \foreach \i in {1,...,\Nsteps} {
            \pgfmathsetmacro{\y}{\x^2 * exp(2.29 - \x) + 0.03}
            \addplot[red, thick, densely dashed] coordinates {(\x,\x) (\x,\y)};
            \addplot[red, thick, densely dashed] coordinates {(\x,\y) (\y,\y)};
            \global\let\x\y
            }

        \addplot[orange, thick] {x^2 * exp(2.10 - x) + 0.03};

        \pgfmathsetmacro{\xstart}{4.29}
        \pgfmathsetmacro{\Nsteps}{1}
        \pgfmathsetmacro{\x}{\xstart}
        \foreach \i in {1,...,\Nsteps} {
            \pgfmathsetmacro{\y}{\x^2 * exp(2.10 - \x) + 0.03}
            \addplot[orange, thick, densely dashed] coordinates {(\x,\x) (\x,\y)};
            \addplot[orange, thick, densely dashed] coordinates {(\x,\y) (\y,\y)};
            \global\let\x\y
            }

        \addplot[purple, thick] {x^2 * exp(1.38 - x) + 0.03};

        \pgfmathsetmacro{\xstart}{2.10}
        \pgfmathsetmacro{\Nsteps}{1}
        \pgfmathsetmacro{\x}{\xstart}
        \foreach \i in {1,...,\Nsteps} {
            \pgfmathsetmacro{\y}{\x^2 * exp(1.38 - \x) + 0.03}
            \addplot[purple, thick, densely dashed] coordinates {(\x,\x) (\x,\y)};
            \addplot[purple, thick, densely dashed] coordinates {(\x,\y) (\y,\y)};
            \global\let\x\y
            }

        \addplot[green, thick] {x^2 * exp(1.13 - x) + 0.03};

        \pgfmathsetmacro{\xstart}{2.17}
        \pgfmathsetmacro{\Nsteps}{1}
        \pgfmathsetmacro{\x}{\xstart}
        \foreach \i in {1,...,\Nsteps} {
            \pgfmathsetmacro{\y}{\x^2 * exp(1.13 - \x) + 0.03}
            \addplot[green, thick, densely dashed] coordinates {(\x,\x) (\x,\y)};
            \addplot[green, thick, densely dashed] coordinates {(\x,\y) (\y,\y)};
            \global\let\x\y
            }

        \addplot[yellow, thick] {x^2 * exp(0.89 - x) + 0.03};

        \pgfmathsetmacro{\xstart}{2.17}
        \pgfmathsetmacro{\Nsteps}{1}
        \pgfmathsetmacro{\x}{\xstart}
        \foreach \i in {1,...,\Nsteps} {
            \pgfmathsetmacro{\y}{\x^2 * exp(0.89 - \x) + 0.03}
            \addplot[yellow, thick, densely dashed] coordinates {(\x,\x) (\x,\y)};
            \addplot[yellow, thick, densely dashed] coordinates {(\x,\y) (\y,\y)};
            \global\let\x\y
            }

        \addplot[red, thick] {x^2 * exp(0.77 - x) + 0.03};

        \pgfmathsetmacro{\xstart}{1.31}
        \pgfmathsetmacro{\Nsteps}{1}
        \pgfmathsetmacro{\x}{\xstart}
        \foreach \i in {1,...,\Nsteps} {
            \pgfmathsetmacro{\y}{\x^2 * exp(0.77 - \x) + 0.03}
            \addplot[red, thick, densely dashed] coordinates {(\x,\x) (\x,\y)};
            \addplot[red, thick, densely dashed] coordinates {(\x,\y) (\y,\y)};
            \global\let\x\y
            }
    \end{axis}
\end{tikzpicture}
    }
    \hfill
    \subfloat[\label{indiv nonchaotic}]{\includegraphics[width=0.49\linewidth]{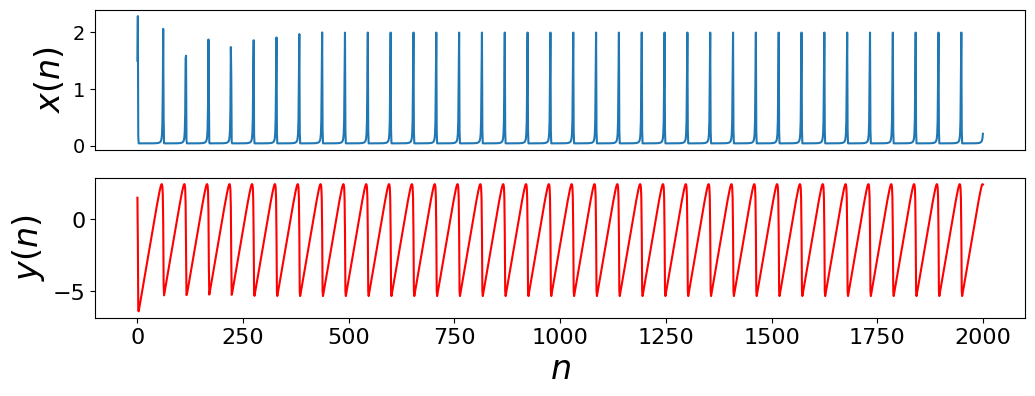}
    }
    \hfill
    \subfloat[\label{indiv chaotic}]{\includegraphics[width=0.49\linewidth]{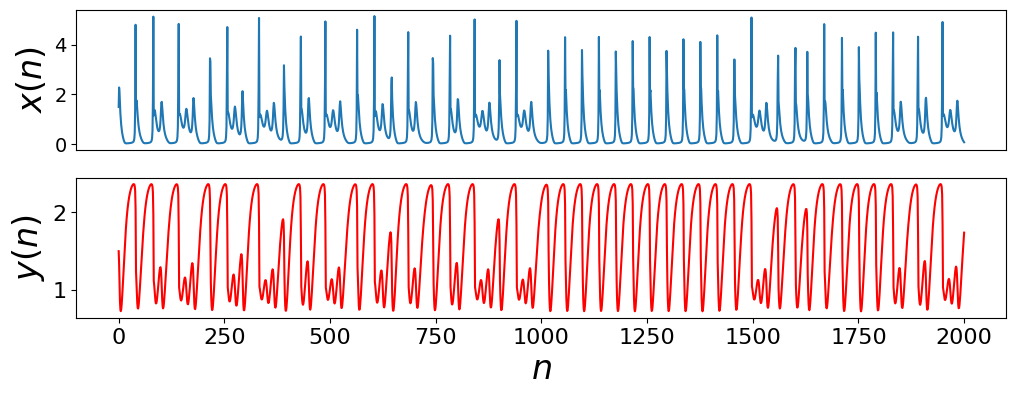}}

    \caption{Different behaviors of the individual Chialvo neuron. (a)~Cobweb plot showing the spiking nonchaotic individual Chialvo neuron $a=1.0$, $b=2.2$, $c=0.26$, and $I=0.04$. The curve shows $x_{n+1}=x_n^2\exp(y-x_n)$ for $y\in \{2.22, 1.51, -0.86, -5.00\}$, with each color signifying a different value of $y$. The dashed line traces the behavior of $x_n$ showing repeated, periodic behavior. (b)~Cobweb plot showing the aperiodic chaotic spiking of the individual Chialvo neuron for $a=0.89$, $b=0.18$, $c=0.28$, and $I=0.03$. The curves plot $x_{n+1}=x_n^2\exp(y-x_n)$ for $y\in \{2.34, 2.29, 2.10, 1.38, 1.13, 0.89, 0.77\}$, with each color signifying a different value of $y$. Contrary to figure (a), there is no pattern and no stable orbit, as evidenced by the lack of structure in the dashed line's trajectory. (c)~$x$ and $y$ versus $n$ for $a=1.0$, $b=2.2$, $c=0.26$, and $I=0.04$ for initial state $(1.5,1.5)$. (d)~$x$ and $y$ versus $n$ for $a=0.89$, $b=0.18$, $c=0.28$, and $I=0.03$ for initial state $(1.5,1.5)$.}
    \label{individual neuron graphs}
\end{figure*}

Since much discussion has previously been given to the influence of the parameters $a$ and $b$ on the system (see Refs.~\cite{kuznetsov,ramirez}), we will consider two sets of parameters $a,b,c,$ and $I$ that have been shown to yield specific behavior. Specifically, the values $a=1.0$, $b=2.2$, $c=0.26$, and $I=0.04$ result in nonchaotic dynamics (Fig.~\ref{indiv nonchaotic}); while $a=0.89$, $b=0.18$, $c=0.28$, and $I=0.03$ result in chaotic dynamics (Fig.~\ref{indiv chaotic}). Unlike the coupled system, one set of parameter values for the individual map often yields the same behavior regardless of initial state. 

Cobweb plots displaying the nonchaotic, periodic behavior and the chaotic, aperiodic behavior are shown in Figs.~\ref{cobweb1} and \ref{cobweb2}, respectively. Figs.~\ref{indiv nonchaotic} and \ref{indiv chaotic} show the resulting nonchaotic and chaotic behavior, respectively, of the neuron over time. The curves in the plane depict $x_{n+1}=x_n^2\exp(y-x_n)+I$ for various values of $y$. As $y$ changes, the curve changes, and the iterative values of $x$ are either pulled upwards or downwards. In the nonchaotic system, the parameters are such that, when $x_n$ is large, $y_{n+1}$ will be small. More concretely, for some $x_n$,
\begin{equation}
    y_{n+1}=1.0\cdot y_n - 2.2\cdot x_n + 0.26,
\end{equation}
which means that as soon as $x_n>0.11$, $y_{n+1}$ will be smaller than $y_n$. Over many iterations, $y$ begins to slowly shrink, eventually inhibiting the spiking of the $x$ variable. In the chaotic system, a large value of $x_n$ does not convey as much influence on $y_{n+1}$, and $x$ is pulled upwards for a longer period of time---again, for some $x_n$,
\begin{equation}
    y_{n+1} = 0.89\cdot y_n - 0.18 \cdot x_n + 0.28.
\end{equation}
Thus, $y_{n+1}>y_{n}$ will only occur if $x_n >-0.6\cdot y_n + 0.28$, which is much more difficult to achieve than the requirement for the nonchaotic system. This causes $x$ to spike higher and more often.

A biological neuron displays a refractory period~\cite{refractory-period}, where, after spiking, it is quiet and does not spike for a period of time. This is achieved by the $y$ variable: when $x$ is near zero, the equation for $y$ becomes
\[y_{n+1}\approx ay_n+c.\]
Therefore, depending on the values of $a$ and $c$, a line will be formed in the $y_n$ versus $y_{n+1}$ plane, and the slope and intercept of this line (and consequently how quickly $y$ reaches a value where the neuron can spike again) are dependent on the chosen values of $a$ and $c$. Thus, $y$ acts as a ``recovery" parameter because of its role in inhibiting the spiking of $x$.

The chaotic system also exhibits metastability, where it spends a period of time in one particular regime of behavior before moving away~\cite{rossi-metastability,kelso}. In Fig.~\ref{indiv chaotic}, the neuron spends approximately 500 iterations in one nearly-periodic behavioral pattern (from 1,000 to 1,500) before being pulled away. It is established that, in order for there to be metastability, there must be coexisting attraction and repulsion~\cite{rossi-metastability}. This simply arises from the parameter values $a,b,c$ and $I$, where the ``attraction" towards a stable orbit arises because, in short, that is the natural tendency of the system. The ``repulsion" comes from the fact that $b<a$, and the $y$ value does not drop sufficiently for $x$ to enter a quiescent refractory period.

\subsection{The coupled system}\label{sec:coupled system}

First, on the construction of the coupled system. Since we are electrically coupling two individual neurons, it is natural to add a coupling term to the injection of current $I$. The coupled neuron system is described by
\begin{align}\label{coupled equations}
    \begin{split}
    \begin{cases}
        x_{n+1} &= x_n^2\exp(y_n-x_n) + I_1 + \Cel_{1,n} \\
        y_{n+1} &= ay_n - bx_n + c \\
        \alpha_{n+1} &= \alpha_n^2\exp(\beta_n-\alpha_n)+I_2+\Cel_{2,n} \\
        \beta_{n+1} &= a\beta_n - b\alpha_n + c,
    \end{cases}
    \end{split}
\end{align}
where $\Cel_{1,n}=g_1(\alpha_n-x_n)$ and $\Cel_{2,n}=g_2(x_n-\alpha_n)$ are the electrical coupling terms, with $g_1=0.05$ and $g_2=0.3$ held constant. The variables $x_n$ and $\alpha_n$ are the voltage variables, while $y_n$ and $\beta_n$ are the recovery variables.

The state space of the coupled system is given by a four-dimensional vector:
\begin{equation}
    \mathbf{x}_n = \begin{pmatrix}
        x_n \\
        y_n \\
        \alpha_n \\
        \beta_n
    \end{pmatrix},
\end{equation}
and the Jacobian matrix of the system is
\begin{widetext}
    \begin{equation}
        \mathbf{J}(\mathbf{x}_n) = 
        \begin{pmatrix}
            2x_n e^{y_n-x_n} - x_n^2 e^{y_n-x_n} - g_1  & x_n^2 e^{y_n-x_n}  & g_1 & 0 \\
            -b & a & 0 & 0 \\
            g_2 & 0 & 2\alpha_n e^{\beta_n-\alpha_n} - \alpha_n^2 e^{\beta_n-\alpha_n} - g_2 & \alpha_n^2 e^{\alpha_n-\beta_n} \\
            0 & 0 & -b & a
        \end{pmatrix}.
    \end{equation}    
\end{widetext}
Then, the QR factorization method of computing Lyapunov spectra~\cite{eckmann,brandon} can be used to characterize the different attractors of the system.

For the set of parameters $a=1.0$, $b=2.2$, $c=0.26$, and $I=0.04$, we found that the individual system is entirely nonchaotic. On the other hand, when two identically nonchaotic neurons with these same parameters are coupled electrically, their interactions cause chaos and aperiodicity. To amplify the small differences in behavior, the identical neurons were coupled asymmetrically; that is, the coupling constants were different: $g_1=0.05$ and $g_2=0.3$. Asymmetry is commonly used in studies of coupled systems to more effectively discriminate between different behaviors~\cite{le,asymmetric-1,asymmetric-2,asymmetric-3}. Notably, the two neurons can interact in a periodic or aperiodic manner, depending entirely on the choice of initial state. As such, a thorough investigation of the different types of behavior is necessary.

First, the case of an initial state $(x_0,y_0,\alpha_0,\beta_0)$ such that $x_0=\alpha_0$ and $y_0=\beta_0$. In this idealistic scenario, $\Cel_1=\Cel_2=0$ because $x_0-\alpha_0=\alpha_0-x_0=0$. Then the equations describe two individual Chialvo neurons with $a=1.0$, $b=2.2$, $c=0.26$, and $I=0.04$; that is, two nonchaotic neurons. These neurons will spike periodically and perfectly in sync, since they are identical, enforcing that $\Cel_{1,2}=0$ for all time. Note, however, that this situation is very unstable and arises only for precisely equal initial conditions. Indeed, no orbit will ever converge upon this pattern of behavior; thus, it is a ``repeller"~\cite{so-repeller}.

Apart from the repeller, the system exhibits two clear regimes of behavior: nonchaotic, periodic bursting; and aperiodic, chaotic bursting. This denotes the system as multistable, meaning there are two distinct, coexisting attractors within one system. We will first briefly analyze the neuronal behavior that these attractors convey, and then we will analyze the basins of attraction that surround the attractors in more detail.

The coupling parameters $\Cel_{1,n}=g_1(\alpha_n-x_n)$ and $\Cel_{2,n}=g_2(x_n-\alpha_n)$ have an important function in the coupled system. The curves in the $x_{n+1}$ versus $x_n$ plane have various fixed points depending on the values of $y$ and $I$. Fig.~\ref{fixed points} shows how, even when $y$ is small, raising the value of $I$ can induce a spike in $x$. For example, if the $y$ value of the system is small or negative, then typically $x$ is quiet and not spiking. However, if the value of $\beta$ is large enough to induce a spike in $\alpha$, then when $\alpha$ spikes, $\Cel_{1}$ also spikes, because $\alpha-x>0$. This, as shown in Fig.~\ref{fixed points}, induces a small spike in the $x$ variable. These interactions are the source of aperiodicity.

\begin{figure}
        \begin{tikzpicture}
    \begin{axis}[
        axis lines=middle,
        xmin=0, xmax=3,
        ymin=0, ymax=3,
        xlabel={$x_n$},
        ylabel={$x_{n+1}$},
        label style = {font=\large},
        xlabel style={below right},
        ylabel style={above left},
        width=8cm,
        height=8cm,
        samples=200,
        domain=0:3,
        legend pos=north east,
        ]
        
        \addplot[
            mark=*,
            only marks,
            mark size=2pt,
            color=green
        ] coordinates {(0.03156, 0.03156)};
        \addplot[
            mark=*,
            only marks,
            mark size=2pt,
            color=blue
        ] coordinates {(1.016, 1.016)};
        \addplot[
            mark=*,
            only marks,
            mark size=2pt,
            color=red
        ] coordinates {(0.71, 0.71)};

        \addplot[black, thick, dashed] {x};
        \addplot[red, thick] {x^2 * exp(0.5 - x) + 0.03};
        \addplot[blue, thick] {x^2 * exp(0.5 - x) + 0.4};
        
        \pgfmathsetmacro{\xstart}{0.03156}
        \pgfmathsetmacro{\Nsteps}{3}
        \pgfmathsetmacro{\x}{\xstart}
        \foreach \i in {1,...,\Nsteps} {
            \pgfmathsetmacro{\y}{\x^2 * exp(0.5 - \x) + 0.4}
            \addplot[blue, thick, densely dashed] coordinates {(\x,\x) (\x,\y)};
            \addplot[blue, thick, densely dashed] coordinates {(\x,\y) (\y,\y)};
            \global\let\x\y
            }

        \pgfmathsetmacro{\xstart}{0.71}
        \pgfmathsetmacro{\Nsteps}{3}
        \pgfmathsetmacro{\x}{\xstart}
        \foreach \i in {1,...,\Nsteps} {
            \pgfmathsetmacro{\y}{\x^2 * exp(0.5 - \x) + 0.03}
            \addplot[red, thick, densely dashed] coordinates {(\x,\x) (\x,\y)};
            \addplot[red, thick, densely dashed] coordinates {(\x,\y) (\y,\y)};
            \global\let\x\y
            }
        \draw[->, thick, blue] (axis cs:1.5,0.9) -- (axis cs:1.5,1.2) node[midway,right]{};
        \draw[->, thick, red] (axis cs:2.0,1.25) -- (axis cs:2.0,1.0) node[midway,right]{};  
    \end{axis}
        \end{tikzpicture}
        \caption{Fixed points and cobweb plot corresponding to $y=0.5$ for a shift from $I=0.03$ (shown in red) to $I=0.4$ (shown in blue). When $\Cel_{1,2}$ spikes, even if $y<1$, the curve will be shifted upwards, inducing a spike. The system begins at the green dot, orbiting according to the red dashed line. When $\Cel_{1,2}$ increases quickly, the curve shifts according to the blue arrow, and the point moves from the location of the red dot to the location of the blue dot. When $\Cel_{1,2}$ goes back to its initial value, the curve follows the red arrow, and the orbit returns to the location of the red dot, following the blue dashed line back to its starting point. In this way, the system is not quiescent even when $y$ is small. This also contributes to aperiodicity in the coupled system.}
        \label{fixed points}
\end{figure}

\begin{figure}
    \centering
    \subfloat[\label{alternate height attractor}]{\includegraphics[width=\linewidth]{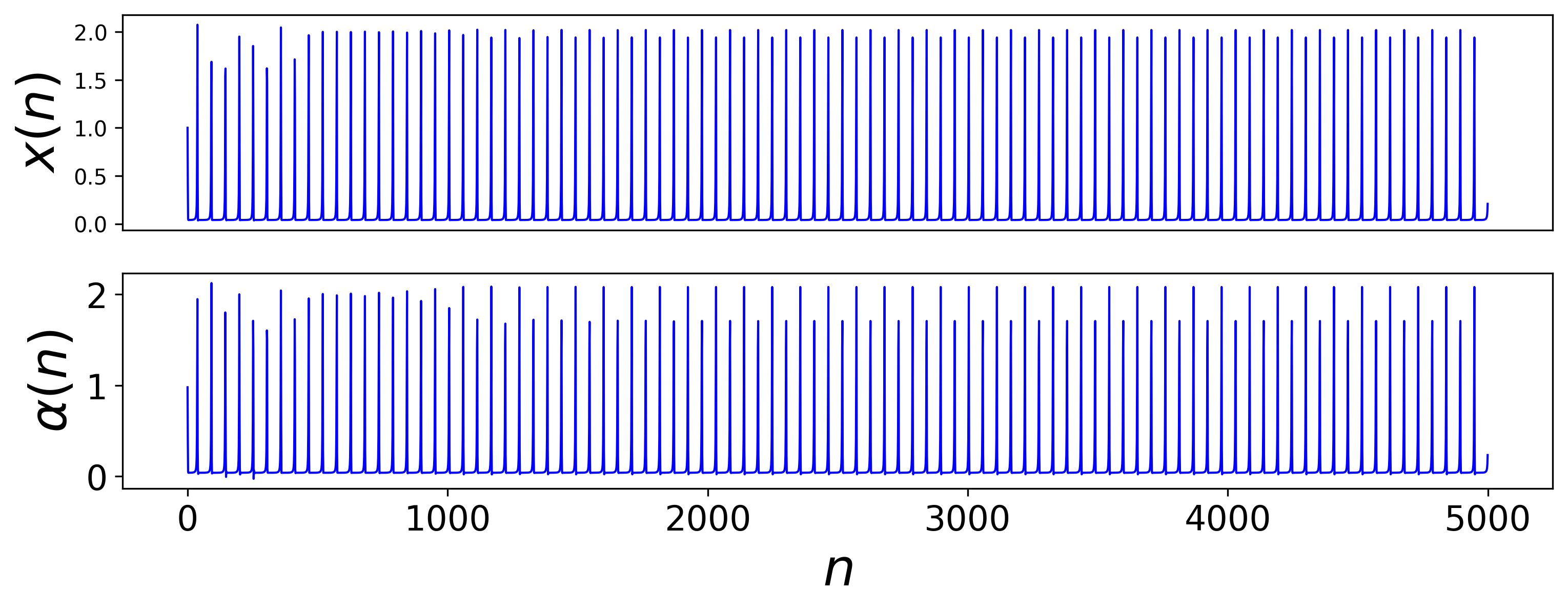}}
    \hfill
    \subfloat[\label{chaotic attractor}]{\includegraphics[width=\linewidth]{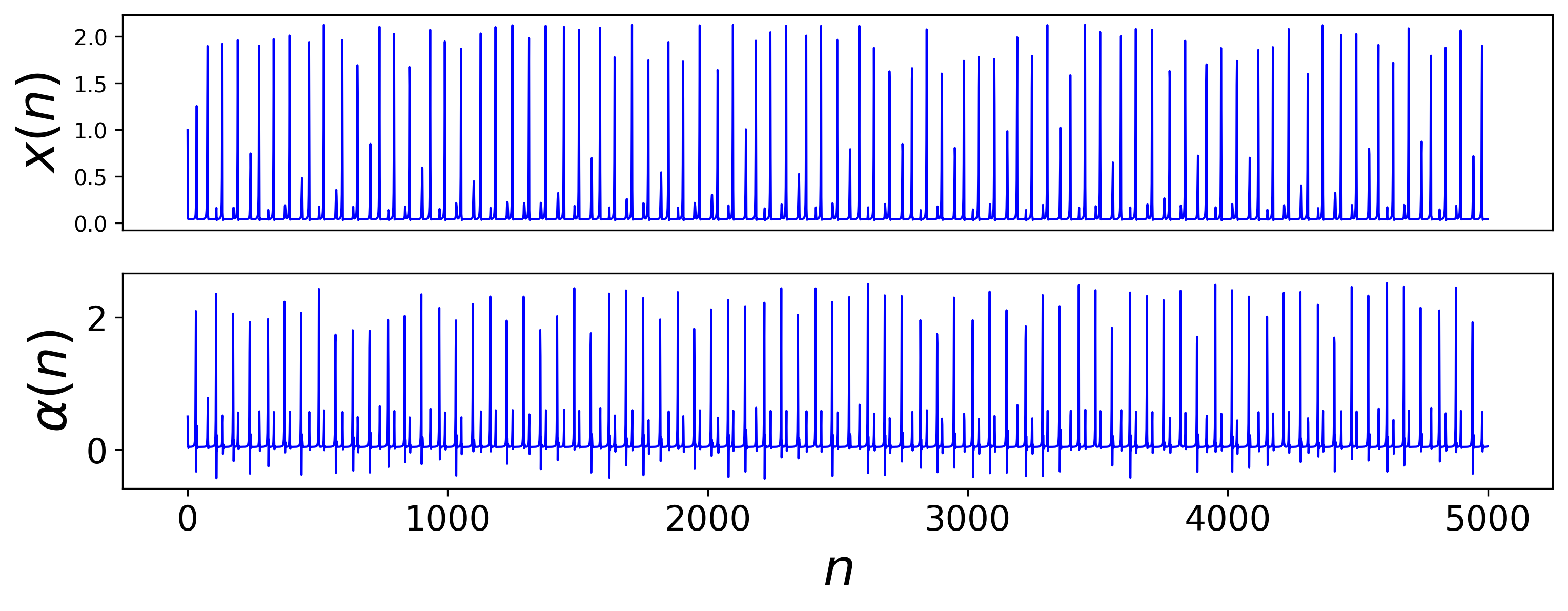}}
    \label{coupled attractors}
    \caption{Behaviors of the voltage variables $x(n),\alpha(n)$ of two coupled Chialvo neurons at the two attractors of the system: (a)~Nonchaotic behavior with initial state $(1.0, 0.0, 0.98, 0.0)$. (b)~Chaotic behavior with initial state $(1.0,0.0,0.5,0.0)$.}
\end{figure}

\begin{figure}
    \centering
    \subfloat[\label{nonchaotic coupling params}]{\includegraphics[width=1.0\linewidth]{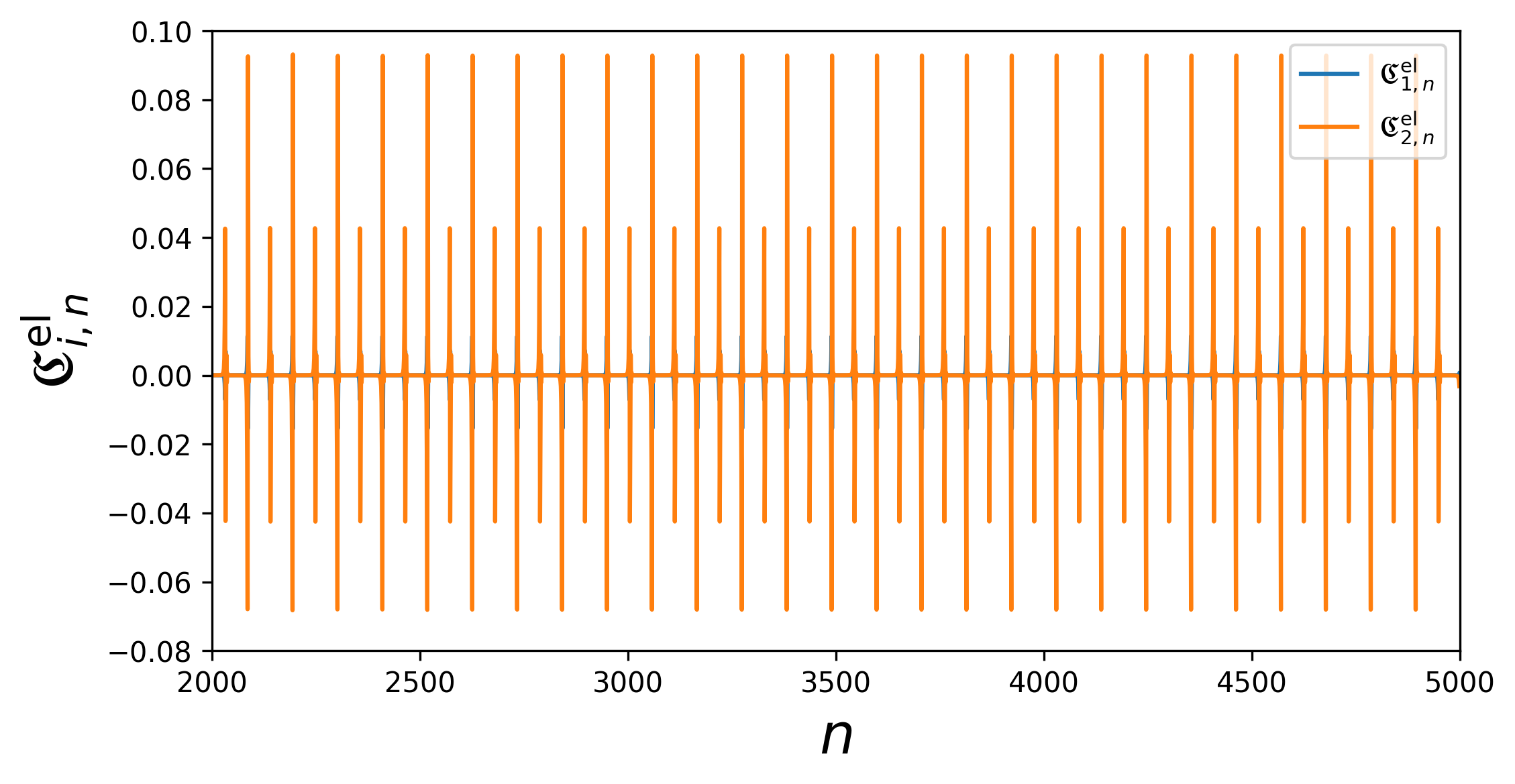}}
    \hfill
    \subfloat[\label{chaotic coupling params}]{\includegraphics[width=1.0\linewidth]{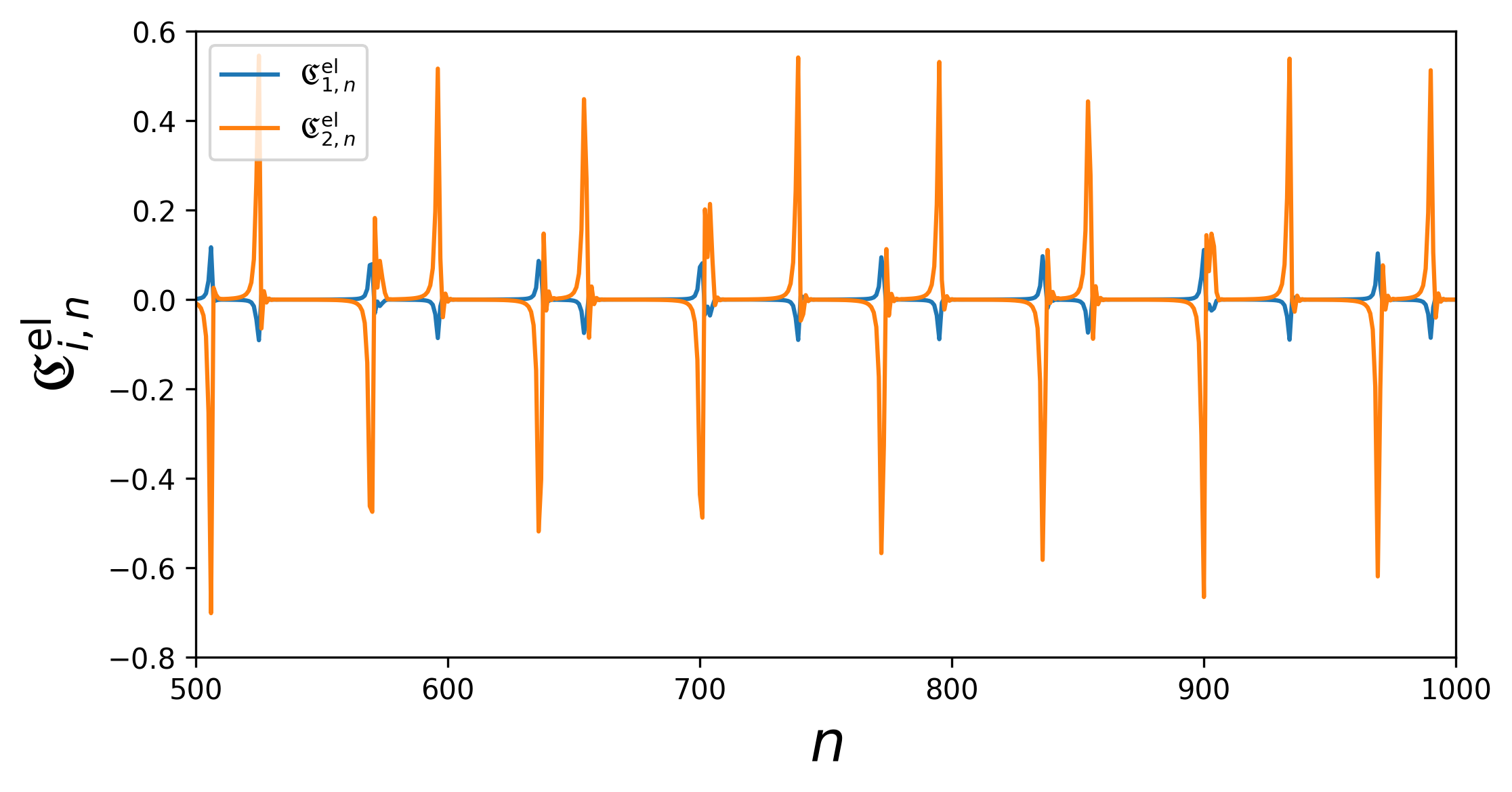}}
    \caption{Behavior of coupling parameters $\Cel_{1,n}$ and $\Cel_{2,n}$ with $a=1.0$, $b=2.2$, $c=0.26$, $I=0.04$. (a)~Nonchaotic, alternate bursting behavior with initial state $(1.0, 0.0, 0.98, 0.0)$. (b)~Chaotic behavior with initial state $(1.0, 0.0, 0.5, 0.0)$.}
    \label{coupling params}
\end{figure}

There is one nonchaotic attractor in this system, and it involves alternate bursting heights. This is due to the asymmetry in the coupling parameters of this system: a spike in one neuron raises the injection of current in the other neuron by an amount dependent on its coupling constant $g$. The voltage variables are shown in Fig.~\ref{alternate height attractor}, while Fig.~\ref{nonchaotic coupling params} shows the coupling parameters alternating heights as the neurons spike. In Fig.~\ref{alternate height attractor}, one may notice a few types of spiking: first, the neuron spikes at random heights, then at equal heights, then it falls into the nonchaotic attractor. This is simply indicative of the route that the orbit takes in state space to reach the attractor. There are multiple ``forces" at play: when one neuron spikes, it changes $I+\Cel$ in the other neuron, causing it to spike to a different height. However, the two neurons spike at almost the exact same time, due to the similarity in $(x_0,y_0)$ and $(\alpha_0,\beta_0)$, causing the system to fall into a stable pattern. The influence of $I+\Cel$ is evident in this stable pattern, as each neuron spikes to an alternating height as $I+\Cel$ changes. 

The chaotic behavior of the system can also be understood mathematically through the coupling parameters. If $(x_0,y_0)$ and $(\alpha_0,\beta_0)$ differ by a large value, then the value of $\Cel$ will cause one of the neurons to spike when the other is quiet (by the same mechanism shown in Fig.~\ref{fixed points}). This causes $\Cel$ in the other neuron to shoot up, and the cycle repeats, causing the neurons to spike back and forth quickly. In this case, the neurons are not spiking at the same time, and spikes occur more frequently. The system never finds a stable pattern of behavior. The voltage variables are shown in Fig.~\ref{chaotic attractor}, and the coupling parameters for the same behavior are shown in Fig.~\ref{chaotic coupling params}.

\begin{figure}
    \centering
    \subfloat[\label{nonchaotic projection}]{\includegraphics[width=\linewidth]{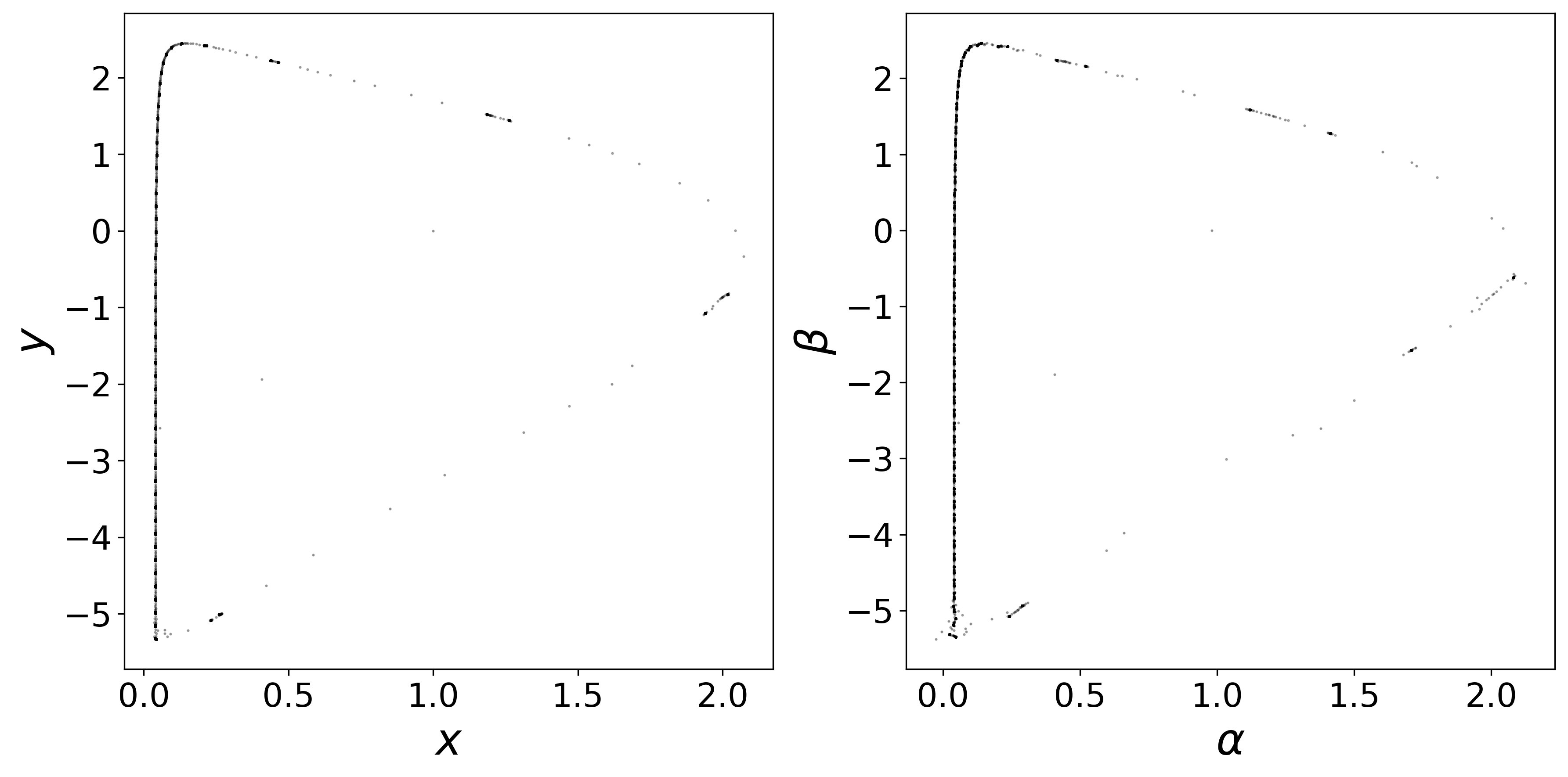}}
    \hfill
    \subfloat[\label{chaotic projection}]{\includegraphics[width=\linewidth]{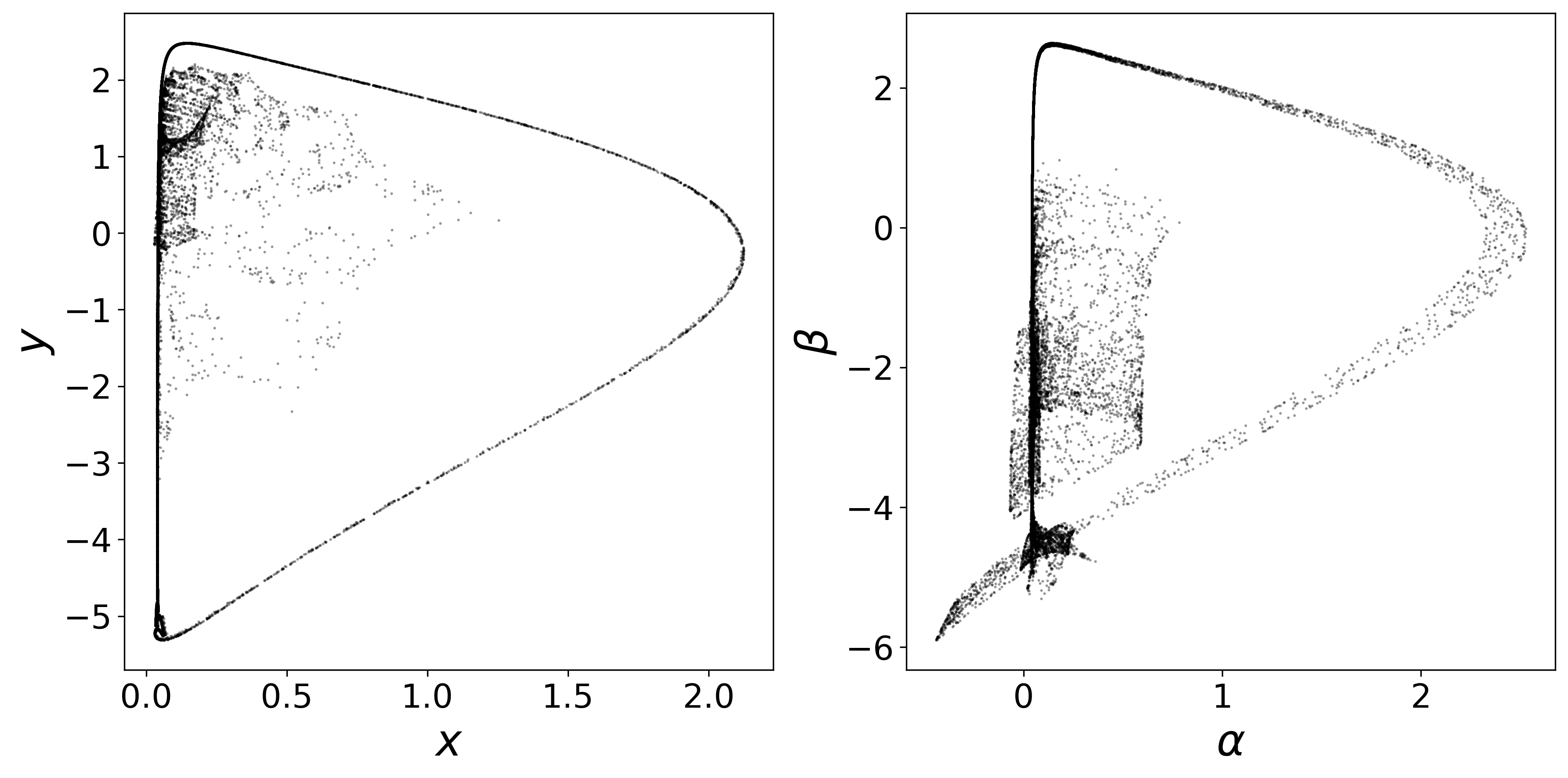}}
    \caption{Projections onto the $xy$ and $\alpha\beta$ planes of orbits of the nonchaotic and chaotic attractors of the coupled Chialvo map with $a=1.0$, $b=2.2$, $c=0.26$, $I=0.04$. (a)~Projections of the nonchaotic attractor from initial state $(1.0, 0, 0.98, 0)$. (b)~Projections of the chaotic attractor from initial state $(1.0, 0, 0.5, 0)$.}
    \label{projections}
\end{figure}

The projection of the attractors themselves onto the $xy$ and $\alpha\beta$ planes is shown in Fig.~\ref{projections}. The difference between the nonchaotic and chaotic regimes is evident: during chaotic bursting, the neurons are asynchronous, and the spike in one neuron causes a low-height spike in the other, which we can see from the clusters of points at small values of $x$ and $\alpha$. Contrarily, the nonchaotic attractor displays clear periodicity and does not deviate from its orbit around the attractor. Furthermore, the projection serves to demonstrate the complicated geometry of the chaotic attractor, especially in contrast with the nonchaotic. This geometry will be more closely studied in later sections.

Before the investigation of the state space, the realism of this system must be mentioned. The purpose of this map is to model a biological neuron; however, issues arise with large initial states in both the voltage variables and recovery variables. Firstly, the issue with large initial states of the recovery variables: since $x_{n+1}\sim\exp(y_n-x_n)$ and $\alpha_{n+1}\sim\exp(\beta_n-\alpha_n)$, a large $y_0$ or $\beta_0$ will cause the voltage variables to spike to unreasonable heights. Then, the large values of the voltage variables will cause the recovery variables, $y$ and $\beta$, to be pulled down to excessively large negative values. $y$ and $\beta$ are limited in how fast they can approach zero from a negative value, and thus it can take many iterations for the orbit to reach the attractor (the attractor is near zero). For example, if $x$ and $\alpha$ are both quiescent at a value near $I_1$ and $I_2$ (that is, 0.04), then $y$ and $\beta$ can be simplified to
\begin{equation}\label{y-beta-rise}
    \begin{cases}
        y_{n+1} &\approx y_n + 0.172 \\
        \beta_{n+1} &\approx \beta_n + 0.172,
    \end{cases}
\end{equation}
which means $y$ and $\beta$ cannot increase by more than $\approx  0.172$ for each iteration, given the current parameter values.

To find the number of iterations it would take $y$ or $\beta$ to reach the attractor, one can divide the distance from 0 by $0.172$. For an example, with initial state $(-10, 50, -30, 20)$, the minimum values of $y$ and $\beta$ are $y_\text{min}\approx -2.4\cdot 10^{28}$, and $\beta_\text{min}\approx -5.6\cdot 10^{27}$. By Eq.~\eqref{y-beta-rise}, the recovery variables $y$ and $\beta$ would take $10^{29}$ and $10^{28}$ iterations, respectively, to reach the attractor and begin their orbit, be it chaotic or nonchaotic. The nuances and results of this fact will be dealt with in Sec.~\ref{sec:basin analysis}.

\section{Basin analysis}\label{sec:basin analysis}

One of the fundamental aspects of this system is its dependence on the initial state. Taking note of the initial states that correspond to certain behaviors builds the state space of the system. By classifying an initial state according to its first Lyapunov exponent $\lambda_1$ (with $\lambda_1>0$ denoting chaotic behavior and $\lambda_1<0$ denoting nonchaotic behavior), the chaotic and nonchaotic basins of attraction can be built in the four-dimensional state space. These basins of attraction will be analyzed as follows. In Sec.~\ref{sec:basin classification}, the stability of the basins is calculated, leading to their classification~\cite{sprott}. In Sec.~\ref{sec:uncertainty exponent}, the uncertainty arising from a fractal basin boundary is quantified. In Sec.~\ref{sec:basin entropy}, the basin entropy is calculated to further quantify the uncertainty in the system and for comparison with the other methods. Finally, in Sec.~\ref{sec:robustness}, the basins are re-analyzed for a variety of different parameter values to ensure that final-state sensitivity and basin structure is an element of the bistable system and not an artifact of our parameter choice.

\subsection{Basin stability and classification}\label{sec:basin classification}

Basin stability provides a concrete way to begin the analysis of the basins of attraction~\cite{basin-stability}. Put simply, for some attractor $A$, the basin stability $\mathcal{B}_A$ of some basin of attraction $\hat{A}$ is the probability that an initial state chosen from a region of state space, denoted $\Omega$, will lie in $\hat{A}$. Given that there are only two attractors in this system, which we will denote $A_c$ (chaotic) and $A_n$ (nonchaotic), the basins should follow that $\mathcal{B}_{A_c} +\mathcal{B}_{A_n} = 1$.

The method of calculation was fairly straightforward. Using the $\Omega$ region given by
\begin{equation}\label{Omega}
    \Omega = \{\mathbf{x} \;|\; x,y,\alpha,\beta \in \left([-2,2]\times [-4,4]\right)^2\},
\end{equation}
a random, uniform distribution of 10,000 initial states in $\Omega$ was sampled and iterated for 20,000 iterations. Denoting chaotic or nonchaotic behavior by the maximal Lyapunov exponent, the number of chaotic and nonchaotic initial states was calculated as a fraction of the total. The basin stabilities were then
\begin{equation}
\begin{split}
    \mathcal{B}_{A_c} &= \frac{\text{\#(chaotic final states)}}{\text{\#(total initial states)}},\\\\
    \mathcal{B}_{A_n} &= \frac{\text{\#(nonchaotic final states)}}{\text{\#(total initial states)}}.
\end{split}
\end{equation}
The stabilities of the chaotic and nonchaotic basins, respectively, were $\mathcal{B}_{A_c} = 0.824$ and $\mathcal{B}_{A_n} = 0.178$ when both were calculated separately. One can see that $\mathcal{B}_{A_c}+\mathcal{B}_{A_n}\approx 1$, and that $\mathcal{B}_{A_c}$ far outweighs $\mathcal{B}_{A_n}$, meaning a random initial state is far more likely to lie in the basin of attraction of the chaotic attractor. 

\begin{figure*}
    \centering
    \subfloat[\label{slice1}]{\includegraphics[width=0.33\linewidth]{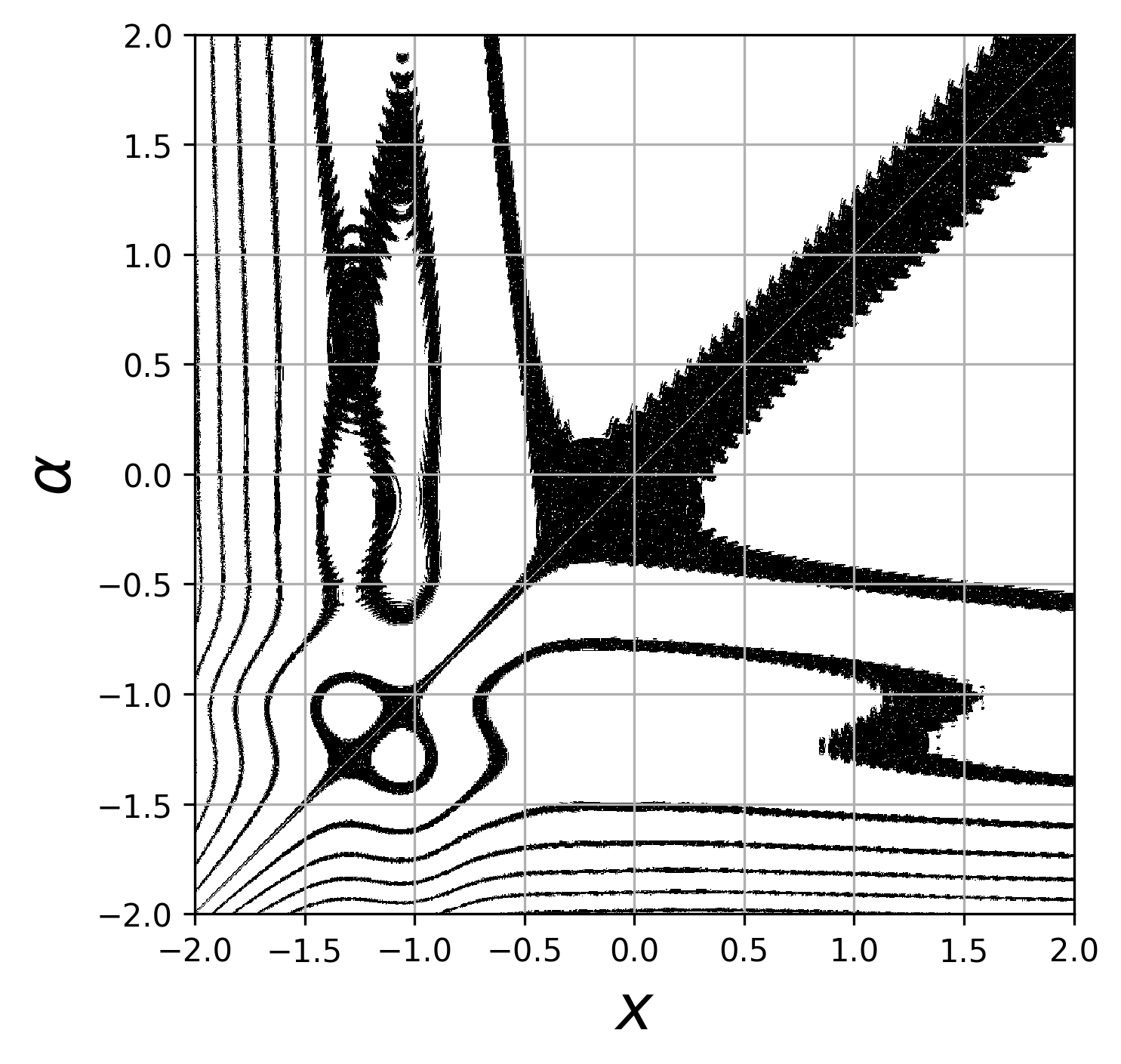}}
\hfill
    \subfloat[\label{slice2}]{\includegraphics[width=0.33\linewidth]{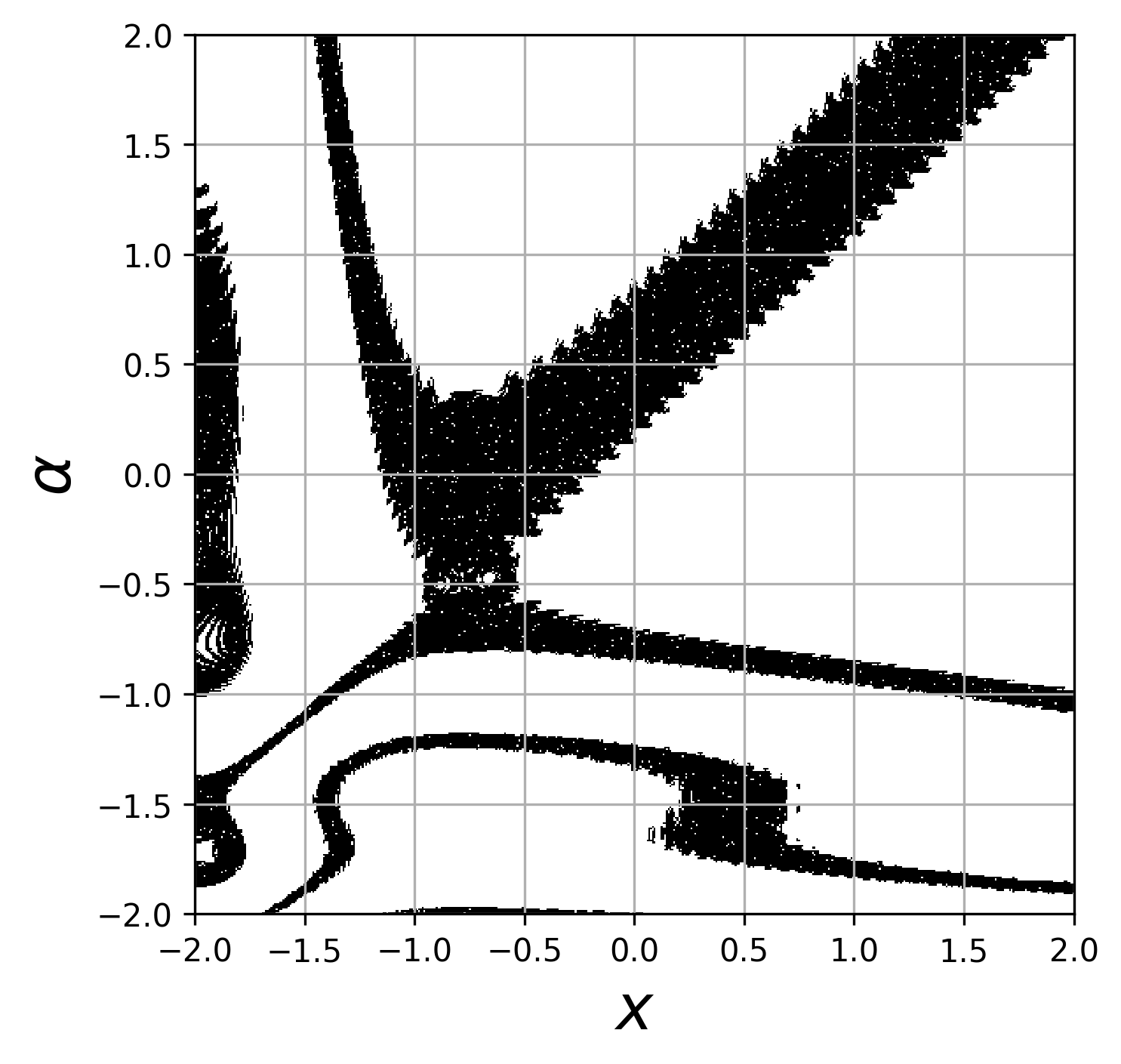}}
\hfill
    \subfloat[\label{slice3}]{\includegraphics[width=0.33\linewidth]{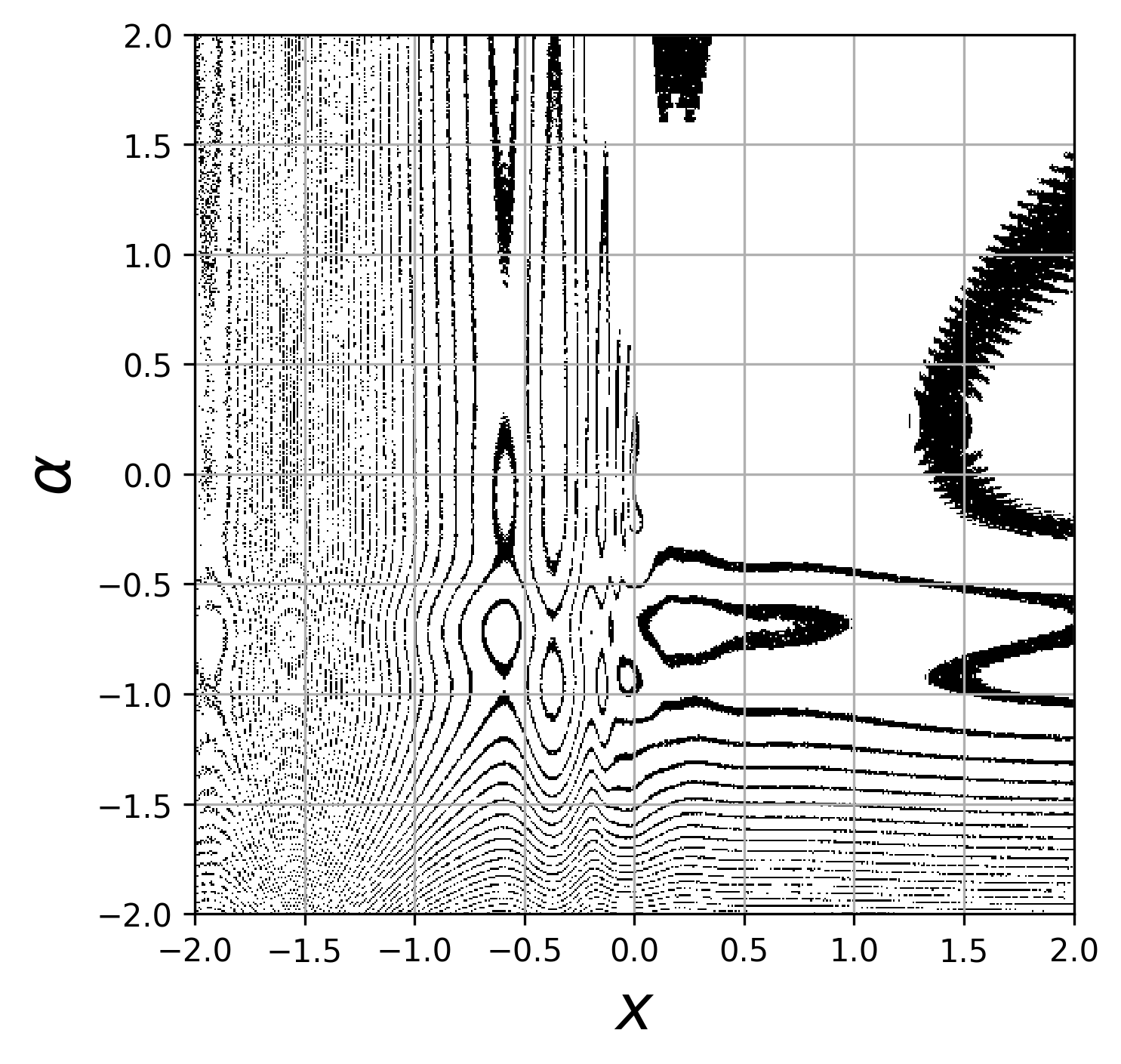}}
    \caption{$x$ vs $\alpha$ slices of the state space of the coupled Chialvo neurons for $-2\leq x,\alpha \leq 2$ with $a=1.0$, $b=2.2$, $c=0.26$, and $I=0.04$. Black denotes the nonchaotic basin and white denotes the chaotic basin. (a) Shows $y=\beta=0$, (b) shows $y=-3.0$, $\beta=-2.0$, (c) shows $y=2.5$, $\beta=1.0$. Note that large, positive values of recovery variables $y$ and $\beta$ are unstable and chaotic or nonchaotic behavior is highly dependent on the location of the voltage variables, hence the disarray in (c). These slices indicate that the basin stabilities $\mathcal{B}_{A_c}$ and $\mathcal{B}_{A_n}$ are near-constant throughout the $\Omega$ region.}
\end{figure*}

These results should correspond roughly to the volumes of state space occupied by each basin. To investigate this, a 2-dimensional slice was taken, holding $y$ and $\beta$ constant at zero, shown in Fig.~\ref{slice1}. The dominance of the chaotic basin is evident through this slice. Since $y$ and $\beta$ were held at equal values, there is a thin white line at the repeller, where $x=\alpha$, as this is where $x=y=\alpha=\beta$, causing equal bursting height. Despite its periodic behavior, since this orbit is enormously unstable and \textit{any} perturbation to one variable will cause it to diverge from the orbit, the maximal Lyapunov exponent remains positive, around $0.005$---thus, the line remains white, denoting a positive maximal Lyapunov exponent. The black basin appears speckled with white spots for this same reason: each initial state spends a different amount of time reaching the nonchaotic attractor. We used 10,000 iterations when taking this slice, and if the neuron took a larger number of iterations to reach the attractor, then the maximal Lyapunov exponent may have stayed positive for longer---occasionally long enough for it to remain positive after all 10,000 iterations were run. If the number of iterations increased, many of the initial states marked as $\lambda_1>0$ would have converged upon $\lambda_1<0$. However, using only 10,000 iterations serves to show that incredibly small changes in initial conditions can yield very different routes to the attractor.

The basin stability analysis was run again for the two-dimensional slice shown in Fig.~\ref{slice1} and obtained results within 10\% for $(\mathcal{B}_{A_c})_2$ and $(\mathcal{B}_{A_n})_2$, indicating that Fig.~\ref{slice1} shows the relative sizes of the basins well in two-dimensional space.

With the context of the stability of the two attractors of the system, the basins can be classified by the method of Sprott and Xiong~\cite{sprott} as follows. Let there be some attractor $A=\{\mathbf{a}_1, \mathbf{a}_2,\dots\}$. Proceed under the assumption that $A$ is infinite, although this method can accommodate attractors with a finite number of points. Define the ``center of mass" of the attractor:
\begin{equation}\label{center of mass}
    \brak{A} = \lim_{N\to\infty}\frac{1}{N}\sum_{i=1}^N\mathbf{a}_i,
\end{equation}
and the standard deviation about the center of mass:
\begin{equation}\label{std center of mass}
    \sigma_A = \sqrt{\lim_{N\to\infty}\frac{1}{N}\sum_{i=1}^N|\mathbf{a}_i - \brak{A}|^2}.
\end{equation}
Then, normalize the distance of some initial condition $\mathbf{x}$ from the attractor as
\begin{equation}\label{normalized distance}
    \xi = \frac{|\mathbf{x} - \brak{A}|}{\sigma_A}
\end{equation}
to normalize the quantification of basins relative to the size of their attractors. 

Next, take an $n$-dimensional ball of radius $\xi$ centered around $\brak{A}$, denoted $S(\xi)$:
\begin{equation}\label{n ball}
    S(\xi) = \{\mathbf{x}:|\mathbf{x}-\brak{A}| < \xi \}
\end{equation}
and observe the fraction of points (from a standardized total number of test points) that fall in the basin of attraction of $A$, denoted $\hat{A}$. Denote $\hat{A}(\xi)$ as the subset of the basin of attraction that is contained in the $n$-ball of radius $\xi$, i.e.
\begin{equation}
    \hat{A}(\xi) = \hat{A}\cap S(\xi)
\end{equation}
then the probability $P(\xi)$ that a particular point in the $n$-ball will lie in the basin of attraction is the ratio of the measures $m(\cdot)$~\cite{measure-theory-tao} of the two sets:
\begin{equation}
    P(\xi) = \frac{m(\hat{A}(\xi))}{m(S(\xi))}.
\end{equation}
Most basins follow a power law~\cite{sprott}:
\begin{equation}\label{p(xi)}
    P(\xi) = \frac{P_0}{\xi^\gamma}
\end{equation}
for some $P_0$ and $\gamma$. It is these $P_0$ and $\gamma$ that serve to classify the basin. The basins can be divided into four categories:
\begin{enumerate}
    \item Class 1 basins include all of state space or all except a set of finite measure and have $P_0=1$ and $\gamma=0$.
    \item Class 2 basins occupy a fixed fraction of state space and have $P_0<1$ and $\gamma=0$.
    \item Class 3 basins extend to infinity but occupy increasingly smaller fractions of state space. These have $0<\gamma<n$ where $n$ is the dimension of the state space.
    \item Class 4 basins occupy a finite region of state space and have a well-defined relative size $\xi_0=P_0^{1/n}$. These basins have $\gamma=n$.
\end{enumerate}
For accuracy, a standard shell method was used~\cite{sprott}, where $\Delta P(2^k)$ is defined to be the probability that a point between the inner radius $\xi=2^k$ and outer radius $\xi=2^{k+1}$ lies in the basin of attraction of $A$. For each $\xi$, $\Delta P(2^k)$ was calculated, and then $P(\xi)$ was found iteratively, using
\begin{equation}\label{iterative shell}
    P(2^{k+1}) = \frac{P(2^k)}{2^n} + \left(1 - \frac{1}{2^n}\right)\Delta P(2^k).
\end{equation}
This ensures that, for each $\xi$, the outer limits of the radius is accurately sampled, eliminating the possibility of using many small radii that have already been measured.

To classify the basins, an iterative Monte Carlo method was used. The center of mass and standard deviation of the attractor were calculated using Eqs.~\eqref{center of mass} and \eqref{std center of mass}, but instead of taking the infinite limit, we simply took $N$ to be some large number of iterations of an orbit that quickly converged upon the chosen attractor. The radius of the hypersphere was exponential:
\begin{equation}
    \xi = 2^k.
\end{equation}
This was repeated for many random values of $\mathbf{x}$ and for $k$ going from 0 to 6, with each orbit run for 20,000 iterations. The iterative shell method given in Eq.~\eqref{iterative shell} was used. To safeguard against extraneous $\lambda_1$ values, two methods were used: firstly, the Jacobian matrices were calculated using only the last 2,000 iterations---when the orbit was already at the attractor. This excluded values of $\lambda_1$ caused by orbits taking a large number of iterations to reach the attractor and thus not indicative of the orbit's final state. Secondly, we rejected quiescent orbits caused by overload by ensuring that the voltage variables were spiking during the last 2,000 iterations. With these checks in place, the number of rejected points rose drastically as the orbit increased. To give context: since 20,000 iterations were used, to avoid being rejected the orbit needed to converge upon the attractor in approximately 18,000 iterations. As shown at the end of Sec.~\ref{sec:coupled system}, the $y$ and $\beta$ values rise at 0.172 per iteration. Thus, the minimum  value that they could take is $-(18,000/0.172) \approx -104,000$. Since $y\sim \exp(y-x)$, in order to keep $y>-104,000$, the system required $(y-x) < 12$. Since, for large $\xi$, the radius being sampled could reach over 100, it is evident that many of the points were going to be rejected. The sampling was run until 5,000 points had been accepted for each attractor; that is, long enough for $P(\xi)$ to converge.

\begin{table}
    \centering
    \caption{Approximate $P(\xi)$ values for the chaotic ($P_{A_c}$) and nonchaotic ($P_{A_n}$) basins, and the probability that a point was rejected for a given value of $\xi$.}
    \begin{ruledtabular}
    \begin{tabular}{cccc}
           $2^k =\xi$  & $P_{A_c}(\xi)$ & $P_{A_n}(\xi)$ & Prob. of Rejection \\
            \hline
           $2^0 = 1$  &  0.828 & 0.178 & 0.0 \\
           
           $2^1 = 2$ & 0.839 & 0.167 & 0.06 \\
           
           $2^2 = 4$ & 0.833 & 0.164 & 0.36 \\
           
           $2^3 = 8$ & 0.846 & 0.163 & 0.59 \\
        
           $2^4 = 16$ & 0.837 & 0.169 & 0.71 \\
        
           $2^5 = 32$ & 0.838 & 0.168 & 0.75  \\
        
           $2^6 = 64$ & 0.837 & 0.155 & 0.78 \\
    \end{tabular}
    \end{ruledtabular}
    \label{p(xi) values}
\end{table}

The Monte Carlo method produced the following results. The nonchaotic basin showed
\begin{equation}
    \ln(P_n(\xi)) = 0.02\ln(\xi) - 3.00,
\end{equation}
yielding $\gamma\approx 0.0$ and $P_0=0.05$, classifying this basin as a Class 2 basin. With error inversely proportional to the number of points sampled, these data had a $\chi^2\approx 0.61$ with a $p$-value of 0.996. 

The chaotic basin showed
\begin{equation}
    \ln(P_c(\xi)) = -0.001\ln(\xi) + 1.43,
\end{equation}
yielding $\gamma\approx 0.0$ and $P_0=0.24$, also signifying a Class 2 basin. For these data, $\chi^2\approx 0.74$ with a $p$-value of $0.998$. 

\begin{figure}
    \centering
    \includegraphics[width=0.9\linewidth]{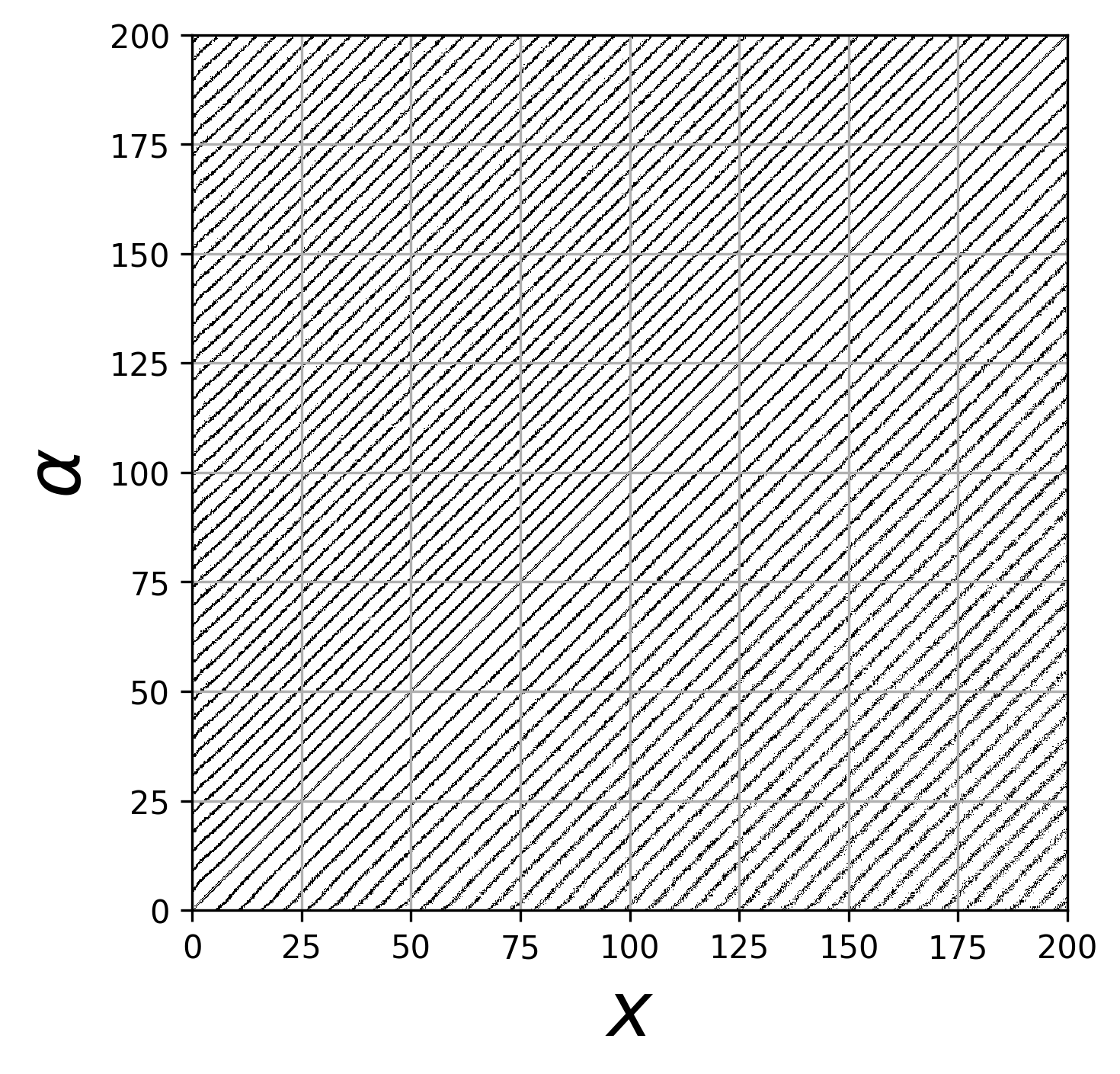}
    \caption{A slice of the coupled Chialvo state space for $\{0\leq x,\alpha \leq 200\}$ with $a=1.0$, $b=2.2$, $c=0.26$, and $I=0.04$.}
    \label{big slice}
\end{figure}

\begin{figure}
    \centering
    \includegraphics[width=0.9\linewidth]{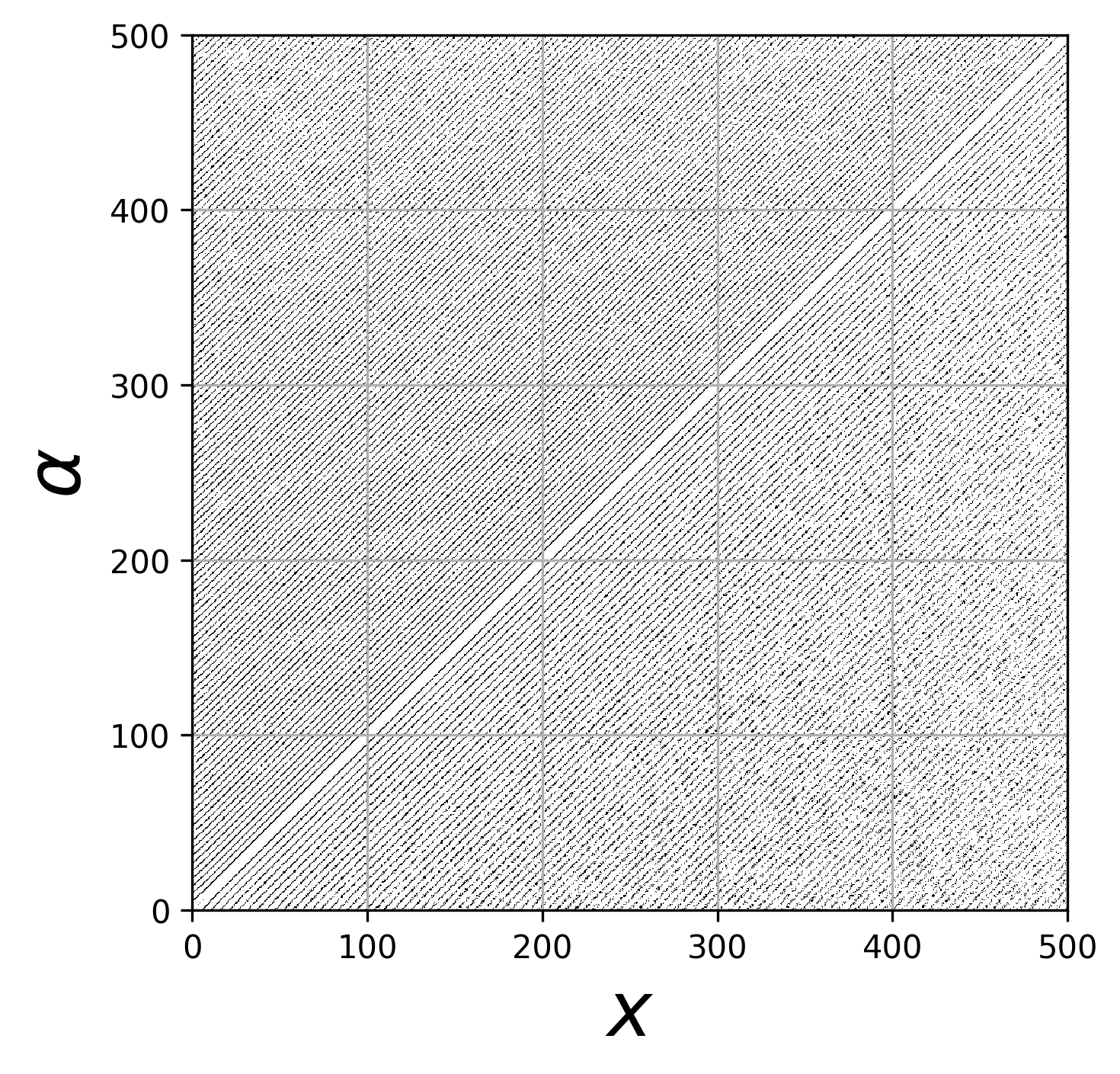}
    \caption{A slice of the coupled Chialvo system for $\{0\leq x,\alpha\leq 500\}$ with $a=1.0$, $b=2.2$, $c=0.26$, and $I=0.04$.}
    \label{big slice 2}
\end{figure}

To further investigate the basins, a large region of the state space was plotted in Figs.~\ref{big slice} and \ref{big slice 2}, holding $y$ and $\beta$ constant at zero to exclude any overload of the system. (Note: for all initial states $(x_0,y_0,\alpha_0,\beta_0)$ where $y_0=\beta_0=0$ and $0\leq x_0,\alpha_0\leq 500$, the orbit converged upon the attractor in less than $7,000$ iterations. The Jacobian matrices were therefore acceptable and $\lambda_1$ was accurate.) The prevalence of both basins is evident, even for large initial conditions $x_0$ and $\alpha_0$. Then the 2D basin stability was calculated for the chaotic attractor $A_c$ in the region $y_0=\beta_0=0$ and $0\leq x_0,\alpha_0\leq 500$. Despite the fact that the nonchaotic basin appears to take up a large portion of this slice, the basin stabilities were $(\mathcal{B}_{A_c})_2=0.861$ and $(\mathcal{B}_{A_n})_2=0.159$. 

When the recovery variables $y$ and $\beta$ are sent to extremely large, negative values, and the orbit is quiescent for a long period of time, it follows that $x\approx\alpha\approx I_1=I_2$. Since the $y$ and $\beta$ variables rise at a much slower rate than the $x$ and $\alpha$ variables, as $y$ and $\beta$ rise, $x\approx\alpha\approx0.04$. Therefore, the system will eventually reach the $\Omega$ region. At this point, since the system is discrete and only the $(n-1)$st iteration has bearing on the $n$th iteration, it is effectively beginning its orbit in the $\Omega$ region, where $\mathcal{B}_{A_c}\approx 0.86$ and $\mathcal{B}_{A_n}\approx 0.14$. This is the basis of the ``chance synchronization" mentioned previously.

This argument also gives context to the horizontal bands of nonchaotic basin that run across the slices of state space in Figs.~\ref{big slice} and \ref{big slice 2}. For these large values of $x$ and $\alpha$, the $y$ and/or $\beta$ will be sent to some large, negative value, causing quiescent behavior in at least one variable. Then, when the system re-enters the $\Omega$ region, it is through chance that the $y$ and $\beta$ variables may be synchronized (causing nonchaotic behavior) or asynchronized (causing chaotic behavior). Indeed, the dominance of the chaotic basin makes sense in this way, too. The $y$ and $\beta$ variables range from approximately $-4$ to $2$. Given that (for $x_n=\alpha_n=0.04$) asynchronous behavior can arise as a result of $|y-\beta| \gtrsim  0.6$, it is not surprising that synchrony is rare in these orbits.

\begin{figure}
    \centering
    \includegraphics[width=1.0\linewidth]{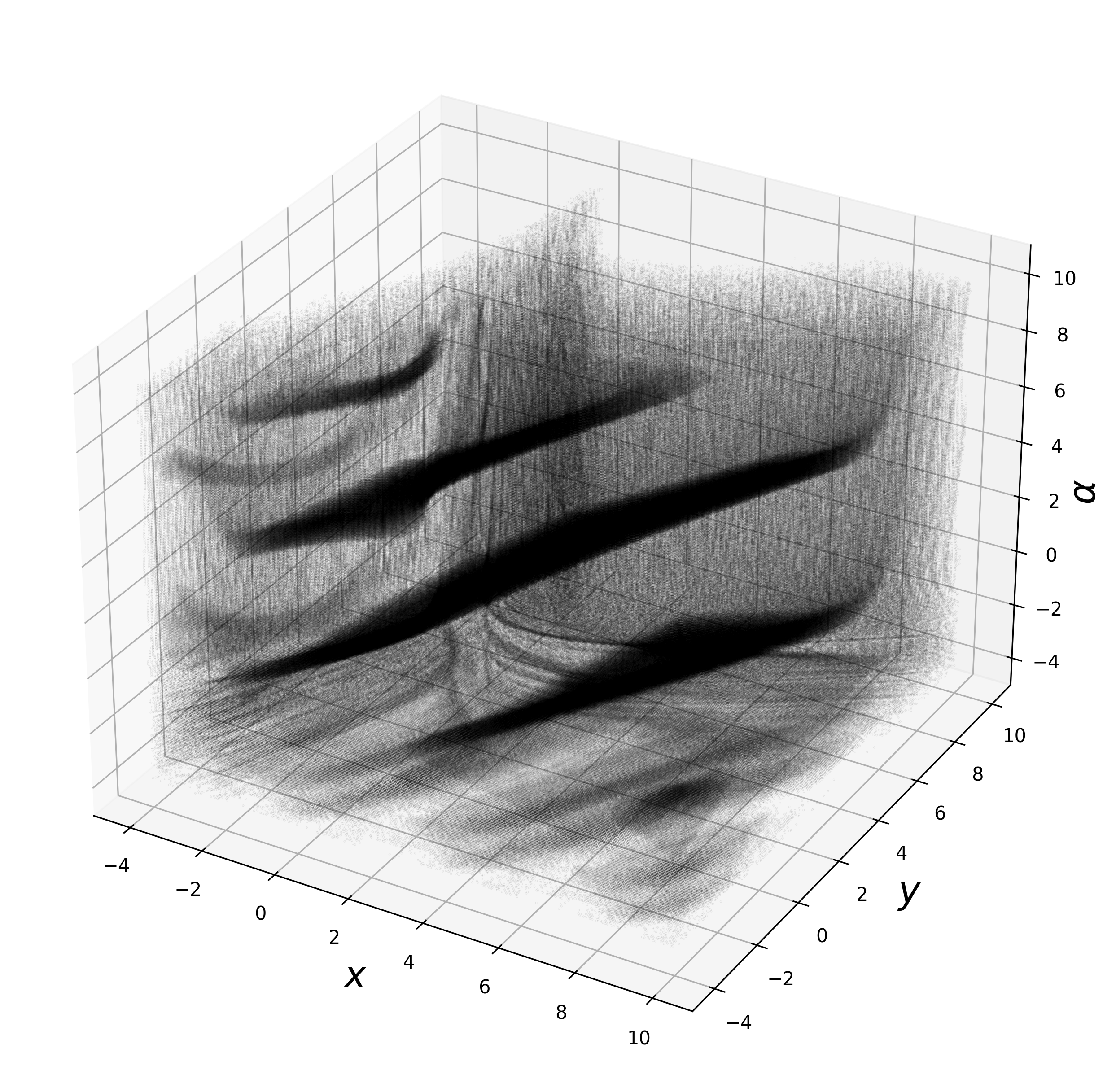}
    \caption{A 3D slice of the coupled Chialvo state space with $\beta=0$ and $-4\leq x,y,\alpha\leq 10$ for $a=1.0$, $b=2.2$, $c=0.26$, and $I=0.04$.}
    \label{fig:3d slice}
\end{figure}

While the behavior of this system at extremes is interesting, it can occlude the results of other analysis methods. However, the following two sections only require analysis of the $\Omega$ region, which is small enough that the extreme behavior rarely emerges and can be handled easily if it does.

\subsection{Uncertainty Exponent}\label{sec:uncertainty exponent}

The variables in a multistable system are often subject to final-state uncertainty, which can manifest as a fractal basin boundary~\cite{grebogi-final-state}. To investigate the presence of a fractal basin boundary, the uncertainty exponent on initial conditions near the basin boundary provides valuable insight~\cite{mcdonald}. Consider a region which contains a basin boundary, and suppose the region is filled with various initial conditions. Then the initial conditions are perturbed by some amount $\epsilon$ (in an arbitrary direction), and consider the fraction of perturbed initial conditions that fall in a different basin than the original point. Denote this fraction $\varrho(\epsilon)$. In a system with smooth basin boundaries, the fraction will go as $\varrho(\epsilon)\sim \epsilon$, but with fractal basin boundaries it will go as
\begin{equation}
\varrho(\epsilon)\sim\epsilon^\mathfrak{u}
\end{equation}
for $\mathfrak{u}\leq 1$ the ``uncertainty exponent." The gravity of this assertion lies in how the reduction of the error $\epsilon$ improves one's ability to predict the final state of a given initial condition. Small values of $\mathfrak{u}$ convey extreme unpredictability of the system, as will be discussed shortly. The fractal dimension~\cite{mandelbrot3} $d$ of the boundary is given by
\begin{equation}\label{fractal dim u}
    d = n-\mathfrak{u}
\end{equation}
where $n$ is the number of dimensions of the entire system. 

For the Chialvo system, a region $\Omega$ of the state space was selected that contained a basin boundary (this was confirmed by taking a slice inside $\Omega$, which is shown in Fig.~\ref{slice1}). The same $\Omega$ region that was given in Eq.~\eqref{Omega} was used:
\[\Omega = \{\mathbf{x} \;|\; x,y,\alpha,\beta \in \left([-2,2]\times [-4,4]\right)^2\},\]
as this region encompasses the attractors themselves, but expands to include the interesting dynamics that result from negative $x$ or $\alpha$ and positive $y$ and $\beta$ that are slightly farther from the attractor. 

\begin{table}
    \centering
    \caption{Approximate $\varrho(\epsilon)$ values.}
    \begin{ruledtabular}
    \begin{tabular}{cc}
        $5\cdot 2^{-k}=\epsilon$ & $\varrho(\epsilon)$ \\
        \hline
        $5\cdot 2^{-1} =$ 2.5 & 0.837 \\
        $5\cdot 2^{-2} =$ 1.25 & 0.821 \\
        $5\cdot 2^{-3} =$ 0.625 & 0.795 \\
        $5\cdot 2^{-4} =$ 0.3125 & 0.712 \\
        $5\cdot 2^{-5} =$ 0.1563 & 0.624 \\
        $5\cdot 2^{-6} =$ 0.0781 & 0.533 \\
        $5\cdot 2^{-7} =$ 0.0361 & 0.500 \\
        $5\cdot 2^{-8} =$ 0.0195 & 0.442 \\
        $5\cdot 2^{-9} =$ 0.00977 & 0.368 \\
    \end{tabular}
    \end{ruledtabular}
    \label{uncertainty exp table}
\end{table}

For this $\Omega$ region, $\epsilon$ was varied, and the percentage of 25 perturbed states that fell into a different basin than the original was noted. In Table \ref{uncertainty exp table}, various values of $\varrho(\epsilon)$ are given and their steady decline is evident. (The value for $k=0$ was rejected because of its large radius.) The data were analyzed and yielded the following results.

A linear regression yielded $\mathfrak{u}=0.182$ with an $R^2$ value of 0.979, indicating good agreement. To put in perspective its title as the ``uncertainty" exponent, this value of $\mathfrak{u}$ tells us that, in order to reduce the uncertainty in which attractor a perturbed state will tend to by a factor of 10, the radius of perturbation $\epsilon$ must be reduced by $10^{5.6}$, since the probability $\varrho(\epsilon)$ scales with $\epsilon^\mathfrak{u}$. Furthermore, Eq.~\eqref{fractal dim u} can now be used to determine that the fractal dimension of the basin boundary is
\begin{equation}\label{boundary fractal dim}
    d = n-\mathfrak{u}\approx 4-0.182 = 3.818.
\end{equation}

This result indicates that the boundary between the two basins is indeed fractal: it is a boundary between two four-dimensional spaces, yet it maintains 3.82 dimensions due to its complicated and fine structure.

\subsection{Basin Entropy}\label{sec:basin entropy}

Another method developed to better describe the nature of an uncertain basin boundary is basin entropy~\cite{basin-entropy}. It aims to quantify the uncertainty that arises when two basins share a fractal boundary, which, as demonstrated above with the uncertainty exponent $\mathfrak{u}$, is the case for the Chialvo system. Basin entropy is calculated as follows. Consider the region $\Omega$, and suppose it contains $N_A$ attractors. Then cover $\Omega$ with boxes of side length $\varepsilon$, and within each box, take various initial states, recording the attractor where they end, and thus giving each initial state a ``color" (two initial states that tend to the same attractor will have the same color). Define $p_{i,j}$ to be the probability that the color $j$ is inside box $i$. From this definition, calculate a value called $S$:
\begin{equation}\label{S}
    S = \sum_{i=1}^N\sum_{j=1}^{m_i}p_{i,j}\log\left(\frac{1}{p_{i,j}}\right)
\end{equation}
where $m_i\in[1,N_A]$ is the number of colors in the $i$th box, and $N$ is the total number of boxes covering $\Omega$. $S$ is, in essence, a sum of the Gibbs entropy of each box. However, since the number of boxes is a constant, $S$ can be normalized to obtain
\begin{equation}\label{basin entropy}
    S_b=\frac{S}{N} = \frac{1}{N}\sum_{i=1}^N\sum_{j=1}^{m_i}p_{i,j}\log\left(\frac{1}{p_{i,j}}\right)
\end{equation}
which is called the basin entropy. This number can range from 0 to $\ln(N_A)$, the latter of which implies complete randomization in the basin structure. 

Assuming that the colors are equiprobable inside any given box, i.e. $p_{i,j}=1/m_i$ for the $i$th box, and that all the basins share one boundary (since this system only has two basins, they must share one boundary), the following manipulations can be made to Eq.~\eqref{basin entropy}. Substituting $p_{i,j}$ for $1/m_i$, it follows that
\begin{equation}
    S_b = \frac{1}{N}\sum_{i=1}^N\sum_{j=1}^{m_i}\frac{1}{m_i}\ln(m_i) = \frac{1}{N}\sum_{i=1}^N\ln(m_i).
\end{equation}
Since $m_i=0$ for all boxes that are not on the basin boundary and $m_i=N_A$ for all boxes that are on the boundary, select from the boundary. 
\begin{equation}
    S_b = \frac{N_b}{N}\ln(N_A)
\end{equation}
where $N_b$ denotes the total number of boundary boxes. This number, $N_b$, scales with the resolution given by the fractal dimension $d$ of the boundary; that is, $N_b\sim \epsilon^{-d}$. $N$, the total number of boxes, simply scales with the dimension of the system, $n$. Thus, $N\sim \epsilon^{-n}$. Then the fraction $N_b/N\sim \epsilon^{-(d-n)}=\epsilon^\mathfrak{u}$ by Eq.~\eqref{fractal dim u}. However, the ratio will have some extra factor, denoted $s$, that corresponds to the size of the boundary and how much of the state space is occupied by it. Finally,
\begin{equation}\label{S_b rewrite}
    S_b = s\ln(N_A)\epsilon^\mathfrak{u}
\end{equation}
which, by varying $\epsilon$, allows us to run the linear regression
\begin{equation}\label{S_b regression}
    \ln(S_b)=\mathfrak{u}\ln(\epsilon)+\ln(s\ln(N_A))
\end{equation}
which we can use to compare with our calculation of the uncertainty exponent $\mathfrak{u}$ to cross-check our results and further analyze the geometries of the different basins. 

\begin{table}[]
    \centering
    \caption{Basin entropy $S_b$ values for different box scales $\epsilon$}
    \begin{ruledtabular}
    \begin{tabular}{cc}
        $5\cdot 2^{-k}=\epsilon$ & $S_b$ \\
        \hline
        $5\cdot 2^{-0}=5$ & 0.3897 \\
        $5\cdot 2^{-1}=2.5$ & 0.3850 \\
        $5\cdot 2^{-2}=1.25$ & 0.3234 \\
        $5\cdot 2^{-3}=0.625$ & 0.3431 \\
        $5\cdot 2^{-4}=0.3125$ & 0.2872 \\
        $5\cdot 2^{-5}=0.1563$ & 0.2824 \\
        $5\cdot 2^{-6}=0.0781$ & 0.2169 \\
        $5\cdot 2^{-7}=0.0391$ & 0.1481 \\
        
    \end{tabular}
    \end{ruledtabular}
    \label{S_b table}
\end{table}

The regression on these data resulted in the equation
\begin{equation}\label{ln(S_b)}
    \ln(S_b)= 0.1760\ln(\epsilon) - 0.8296
\end{equation}
with an $R^2$ value of 0.839. The slope of this line shows agreement with the uncertainty exponent of $0.182$ from the previous section. Note that the points in Table \ref{S_b table} do not descend exactly with some slope, but some fluctuations are expected simply due to chance. This agreement serves to further solidify the findings of Sec.~\ref{sec:uncertainty exponent}.

The $y$-intercept of the regression can also provide information. From Eq.~\eqref{S_b regression}, note that the $y$-intercept is of the form $\ln(s(\ln(N_A))$. Conveniently, $N_A=2$. By simple algebra, $s=0.63$. This constant serves to normalize Eq.~\eqref{S_b rewrite}, is independent of $\epsilon$, and is related to the concept of lacunarity~\cite{mandelbrot2}. Lacunarity describes a fractal's deviation from translational invariance and connectedness~\cite{daza-entropy}, but $s$ is not exactly the same. It encodes how much the basins become ``mixed" inside each box as $\epsilon$ changes. A small value of $s$ indicates that, as $\epsilon$ grows, the basin content of the boxes is not changing. Contrarily, for a dense and uncertain area, a small change in $\epsilon$ may greatly change the number, position, or quantity of each basin inside each box, and $s$ may approach 1 or even exceed it.

Our $s$ value of 0.63 indicates that the boundary is indeed fractal, although the basins are not so densely interspersed as to cause an $s$ value larger than 1; this may occur, for example, in a riddled basin. This result could be expected: the uncertainty exponent indicates that the basin was fractal, but by looking at the slices of the phase space it is evident that there are also clear, separated areas of only nonchaotic or only chaotic behavior. Thus, as the resolution $\epsilon$ is increased, the ``mixing" of the basins inside each box is not extremely high. Conversely, the basins are not disconnected, so an extremely small $s$ value would also not be expected.

\subsection{Robustness of Results}\label{sec:robustness}

\begin{figure*}
    \centering
    \includegraphics[width=0.75\linewidth]{Images/uncertainty_exponent_sweep.png}
    \caption{Uncertainty exponent $\mathfrak{u}$ for $c=0.26,I=0.04,g_1=0.05,g_2=0.3$. The points $(a,b)$ resulting in a quiescent or chaotic individual neuron were discarded and are shown in gray. The points yielding nonchaotic individual units displayed final state sensitivity when coupled; some contained a chaotic pseudoattractor. The phenomenon of final state uncertainty is therefore not confined strictly to our chosen parameter set, but it is also not a property of every choice of parameters for the coupled Chialvo system.}
    \label{fig:uncertainty exp sweep}
\end{figure*}

\begin{table*}
    \centering
    \caption{Results after a parameter sweep of $a,b,c,I,g_1,g_2$ for the parameter sets that resulted in nonchaotic individual neurons where the coupled system retained both the chaotic and nonchaotic basins. The quasimultistable systems displayed a pseudoattractor that remained for up to 30000 iterations; however, in order to accurately capture its structure, time steps were limited to the lifetime of the pseudoattractor when extracting $\mathfrak{u}$. Final state uncertainty is present in all cases.}
    \begin{ruledtabular}
    \begin{tabular}{ccccc}
        & $(a,b,c,I,g_1,g_2)$ & $\mathfrak{u}$ & $\mathcal{B}_{A_c},\mathcal{B}_{A_n}$ & Basin Class ($c/n$) \\
        \hline
        & $(0.95, \quad 0.95, \quad 0.28,\quad  0.04, \quad 0.02,\quad  0.05)$ & $0.003 $ & $0.57/0.43$ & 2/2 \\
        Quasimultistable & $(0.96,\quad 0.55,\quad 0.28,\quad 0.04,\quad 0.02,\quad 0.05)$ & $0.001$ & $0.41/0.59$ & 2/2 \\
        & $(0.97,\quad 0.65,\quad 0.28,\quad 0.04,\quad 0.02,\quad 0.30)$ & $0.004$ & $0.55/0.45$ & 2/2 \\
        & $(0.97,\quad 1.25,\quad 0.28,\quad 0.04,\quad 0.02,\quad 0.02)$ & $0.003$ & $0.41/0.59$ & 2/2 \\
        \hline
        & $(0.94,\quad 0.85,\quad 0.28,\quad 0.04,\quad 0.02,\quad 0.05)$ & $0.086$ & $0.57/0.43$ & 2/2 \\
        & $(0.96,\quad 1.30,\quad 0.28,\quad 0.04,\quad 0.02,\quad 0.20)$ & $0.043$ & $0.55/0.45$ & 2/2 \\
        Multistable & $(0.98,\quad 1.90,\quad 0.28,\quad 0.04,\quad 0.05,\quad 0.20)$ & $0.457$ & $0.48/0.52$ & 2/2 \\
        & $(0.99,\quad 2.10,\quad 0.28,\quad 0.04,\quad 0.02,\quad 0.05)$ & $0.016$ & $0.54/0.46$ & 2/2 \\
        & $(0.99,\quad 2.20,\quad 0.28,\quad 0.04,\quad 0.20,\quad 0.05)$ & $0.555$ & $0.61/0.39$ & 2/2 \\
        & $(1.00,\quad 2.20,\quad 0.28,\quad 0.04,\quad 0.05,\quad 0.05)$ & $0.039$ & $0.23/0.77$ & 2/2
    \end{tabular}
    \end{ruledtabular}
    \label{tab:parameter sweep}
\end{table*}

To ensure that these results were not confined to the specific parameter set given, a sweep was run across the variables $a,b,c,I,g_1,$ and $g_2$, and the numerics were repeated. As shown in Ref.~\cite{kuznetsov}, the $a$-$b$ parameter space of the individual neuron can be divided into a periodic region, P, two-frequency quasi-periodic region, 2T, and a chaotic region, C. To ensure chaoticity and basin structure arose from coupling and not from invididual units, parameter sets that produced chaotic individual neurons were discarded. Then, the P and 2T regions were scanned. Below the first Neimark-Sacker (NS) bifurcation at low $b$ and past the second NS bifurcation at high $b$, the fixed point is globally attracting and the system is monostable; therefore, the coupled neurons did not produce a bistable system. Between the two bifurcations, the P, 2T, and C regions are interspersed, and interaction dynamics arise. It was found that bistable systems from nonchaotic individual neurons occurred in the P region between the two NS bifurcations, where coupling can draw the orbit from periodicity into chaos or quasi-periodicity. The parameter set discussed in this paper lies on one of the periodic Arnold resonant tongues between the chaotic and quasi-periodic regions. 

Shown in Fig.~\ref{fig:uncertainty exp sweep}, a coarse $a$-$b$ parameter space colormap was obtained for the uncertainty exponent $\mathfrak{u}$, with the $a$ and $b$ ranges chosen to encompass the area of the $a$-$b$ plane with the densest prevalence of bistability. The points that were discarded because the isolated unit had no dynamics or had strictly chaotic dynamics are shown in gray. The points that were nonchaotic individually produced a bistable system with final state sensitivity. These points are colored, and the magnitude of $\mathfrak{u}$ is shown in the colorbar. Moreover, in many cases the system displayed quasimultistability \cite{le}, where a state would remain in the orbit of a chaotic pseudoattractor for up to 30000 iterations before collapsing back to periodic behavior. For sufficiently low time steps, these systems displayed extreme final-state sensitivity, with $\mathfrak{u}\approx 0$. Parameters yielding quasimultistability are enclosed in boxes in Fig.~\ref{fig:uncertainty exp sweep}. 

For representative examples, a collection of $(a,b)$ values were chosen randomly between the two NS bifurcations for $c=0.28,I=0.04$. As before, quiescent or chaotic individual units were discarded. The results are shown in Table~\ref{tab:parameter sweep}, where quasimultistability with $\mathfrak{u}\approx 0$ again emerges, with the basins occupying approximately equal portions of state space. The multistable region, from which the main set of parameters studied in this paper emerged, also displayed the expected behavior, with $\mathfrak{u}<0.6$ for all cases, and two class-2 basins occupying fixed portions of state space.

We can conclude that the presence of final-state uncertainty is not limited to our chosen set of parameter values, but that there are regimes which will not give rise to bistability in the coupled system. Of the parameter sets that gave rise to bistable or quasi-bistable systems, all of them exhibited final-state uncertainty with fractal basin boundaries.

\section{Conclusions and Future Work}\label{sec:conclusions}
This study observed rich behavior resulting from the electrically coupled Chialvo neuron system. The basins of attraction were analyzed, classified, and quantified using various methods, including basin stability, basin classification, basin entropy, and uncertainty exponents. The system was found to contain one chaotic attractor and one nonchaotic attractor. The chaotic attractor occupies the vast majority of the state space of the system---around 86\%---in both the four-dimensional $\Omega$ region and in two-dimensional slices of the voltage variables $x$ and $\alpha$. The nonchaotic basin takes up the remaining space, around $14\%$, extending to infinity in thin lines through a form of chance synchronization of the recovery variables $y$ and $\beta$. It was concluded that the basins are both Class 2 basins, occupying fixed fractions of the state space. Analysis of slices of the state space and the basin stability for increasing radius agrees with these findings.

The boundary between the two basins is fractal and has a fractal dimension of approximately $3.8$, indicating that although it separates two four-dimensional spaces, its fractal structure causes it to have the properties of a four-dimensional object. The uncertainty exponent reveals that in order to improve final state uncertainty by a factor of 10, one needs to improve the initial state uncertainty by a factor of $10^{5.6}$. Finally, the basin entropy of the $\Omega$ region showed agreement with the calculated uncertainty exponent and provided insight into the fractal nature of the basin.

We believe these methods can be applied to other discrete-time or continuous-time neuronal models and hope to see further studies conducted on the subject. As shown in Ref.~\cite{ubiquity}, we believe that final-state sensitivity is ubiquitous in discrete-time systems, and a closer look at the mechanisms behind each model may provide context for the others. The application to continuous-time systems is a clear next step, and while we believe continuous time systems may exhibit these same properties, further research is needed. While quasimultistability~\cite{le} is not strictly present in this system, we note that the neurons may exhibit various behaviors before converging upon a given attractor. Thus, we suggest further studies into the possible existence of quasimultistability, given that transient synchronization has already been found in a Chialvo system~\cite{chialvo-memristor}.

These findings may have some biological implications. It has been shown that memory and neuronal communication are reliant on the synchronous firing of neurons~\cite{fries,fell-coupling,fell-phase-synchronization}. The dysregulation of synchrony in neurons can lead to neurological diseases like Parkinson's disease~\cite{parkinson,parkinson-2}, Alzheimer's disease~\cite{alzheimer,alzheimer-2}, or epilepsy~\cite{epilepsy-1,epilepsy-2,epilepsy-3}. This dysregulation involves large networks of neurons that are coupled in ways not explored by this paper; thus, the translation of this model's behavior to large networks with various coupling mechanisms could be explored.

Multistability has also been observed experimentally~\cite{lee,heyward,loewenstein}, and thus minute differences in the state of a biological neuron may produce vastly different outcomes. We suggest building upon experimental studies~\cite{elson,abarbanel,varona,guo,an} to investigate links between the behavior outlined in this paper and behavior in real biological neurons.

\bibliography{refs}

\end{document}